\documentclass[prd,twocolumn,amsmath,amssymb,floatfix,nofootinbib,superscriptaddress]{revtex4}
\usepackage{graphicx}
\usepackage{bm}
\usepackage{amsmath,amssymb}
\usepackage{epsfig}
\usepackage{color}
\usepackage{natbib}
\usepackage{enumerate}
\usepackage{ifthen}
\usepackage{subfigure}
\usepackage{hhline}
\usepackage{multirow}
\newcommand{\planck}{\textit{Planck}}

\newcommand{\nueff}{\nu_{\rm eff}}
\newcommand{\GHz}{{\rm GHz}}
\newcommand{\Ghz}{\GHz}

\newcommand{\fsky}{f_{\rm sky}}
\newcommand{\vepsilon}{\mathbf{\epsilon}}
\newcommand{\CORE}{CORE}
\newcommand{\PRISM}{PRISM}
\newcommand{\PIXIE}{PIXIE}

\newcommand{\Planck}{\textit{Planck}}

\newcommand{\nH}{n_{\rm H}}
\newcommand{\nHe}{n_{\rm He}}

\newcommand{\lmax}{l_{\text{max}}}

\providecommand{\CAMB}{\textsc{camb}}

\newcommand{\begm}{\begin{pmatrix}}
\newcommand{\enm}{\end{pmatrix}}

\newcommand\ba{\begin{eqnarray}}
\newcommand\ea{\end{eqnarray}}
\newcommand\bea{\begin{eqnarray}}
\newcommand\eea{\end{eqnarray}}

\newcommand\be{\begin{equation}}
\newcommand\ee{\end{equation}}

\providecommand{\Tr}{\text{Tr}}
\newcommand{\la}{\langle}
\newcommand{\ra}{\rangle}

\newcommand{\mC}{\bm{C}}

\newcommand{\mN}{\bm{N}}

\newcommand{\mR}{\bm{R}}

\newcommand{\boldvec}[1]{{{\mathbf{#1}}}}

\newcommand{\vT}{\boldvec{T}}

\newcommand{\vn}{\boldvec{n}}

\newcommand{\vv}{\boldvec{v}}

\newcommand{\vx}{\boldvec{x}}

\newcommand{\clo}{\mathcal{O}}

\newcommand{\vnhat}{\hat{\vn}}

\def\eprinttmp@#1arXiv:#2 [#3]#4@{
\ifthenelse{\equal{#3}{x}}{\href{http://arxiv.org/abs/#1}{#1}}{\href{http://arxiv.org/abs/#2}{arXiv:#2} [#3]}}

\providecommand{\eprint}[1]{\eprinttmp@#1arXiv: [x]@}
\newcommand{\adsurl}[1]{\href{#1}{ADS}}
\providecommand{\bibinfo}[2]{\ifthenelse{\equal{#1}{isbn}}{
\href{http://cosmologist.info/ISBN/#2}{#2}}{#2}}

\begin{document}

\title{Rayleigh scattering: blue sky thinking for future CMB observations}

\author{Antony Lewis}
\homepage{http://cosmologist.info}
\affiliation{Department of Physics \& Astronomy, University of Sussex, Brighton BN1 9QH, UK}

%\date{\today}

\begin{abstract}
Rayleigh scattering from neutral hydrogen during and shortly after recombination causes the CMB anisotropies to be significantly frequency dependent
at high frequencies. This may be detectable with \Planck, and would be a strong signal in any future space-based CMB missions.
The later peak of the Rayleigh visibility compared to Thomson scattering gives an increased large-scale CMB polarization signal that is a greater than $4\%$ effect for observed frequencies $\nu \agt 500\GHz$. There is a similar magnitude suppression on small scales from additional damping. Due to strong correlation between the Rayleigh and primary signal, measurement of the Rayleigh component is limited by noise and foregrounds, not cosmic variance of the primary CMB, and should observable over a wide range of angular scales at frequencies $200 \GHz\alt \nu \alt 800 \GHz$.
I give new numerical calculations of the temperature and polarization power spectra, and show that future CMB missions could measure the temperature Rayleigh cross-spectrum at high precision, detect the polarization from Rayleigh scattering, and also accurately determine the cross-spectra between the Rayleigh temperature signal and primary polarization. The Rayleigh scattering signal may provide a powerful consistency check on recombination physics. In principle it can be used to measure additional horizon-scale primordial perturbation modes at recombination, and distinguish a significant tensor mode $B$-polarization signal from gravitational lensing at the power spectrum level.
\end{abstract}
\maketitle

%\vskip .2in\
\section{Introduction}

 Neutral hydrogen produced as the universe recombined at redshift $z\sim 1000 $ is often modelled as being transparent, so that photons only scatter from residual free electrons. However neutral hydrogen can also interact with and scatter radiation. Since recombination only happens once the typical photon energy is well below the ionization energy, by the time hydrogen is produced almost all photons will have wavelengths much larger than the atomic radius.  The classical scattering of long-wavelength photons from the dipole induced in the neutral hydrogen is then called Rayleigh scattering, which has an asymptotic $\nu^4$ scaling with frequency. Higher frequencies of the observed CMB anisotropies will therefore be Rayleigh scattered during and shortly after recombination.

 On small, sub-horizon scales Rayleigh scattering  leads to a damping of the anisotropies as photons from hot spots are scattered out of the line of sight, and photons from cold spots are mixed with photons scattering into the line of sight. Rayleigh scattering therefore gives the observed small-scale CMB hot spots a red tinge, for the same reason that sunsets look red. The small-scale polarization signal is also reduced by the additional scattering for a similar reason. However
 the large-scale polarization from recombination is due to coherent quadrupole scattering \emph{into} the line of sight. The additional Rayleigh scattering at late times, where the quadrupole is larger,
 therefore increases the polarization at high frequencies,
 so the polarized sky is slightly blue on large scales.

\begin{figure}
\begin{center}
\vspace{1cm}
\includegraphics[width=8cm]{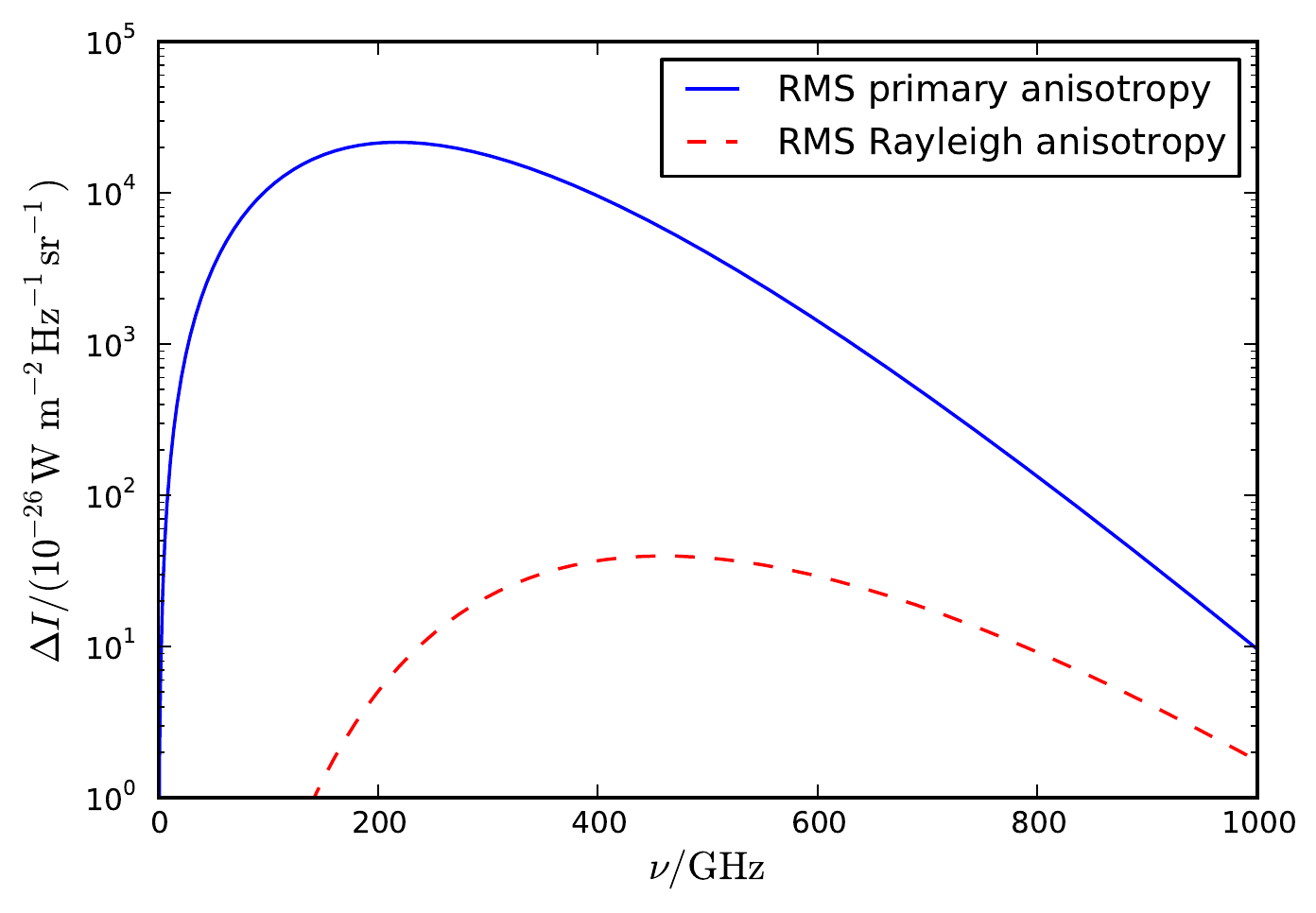}
\caption{Brightness intensity of the root mean square (RMS) CMB temperature anisotropy at $l\le 2000$ as a function of frequency, for the primary signal (no Rayleigh scattering, solid line) and the Rayleigh scattering contribution (scaling with a relative factor approximately proportional to $\nu^4$, dashed line).
\label{rms_spectra}
}
\end{center}
\end{figure}

\begin{figure*}
\begin{center}
\includegraphics[width=8.8cm]{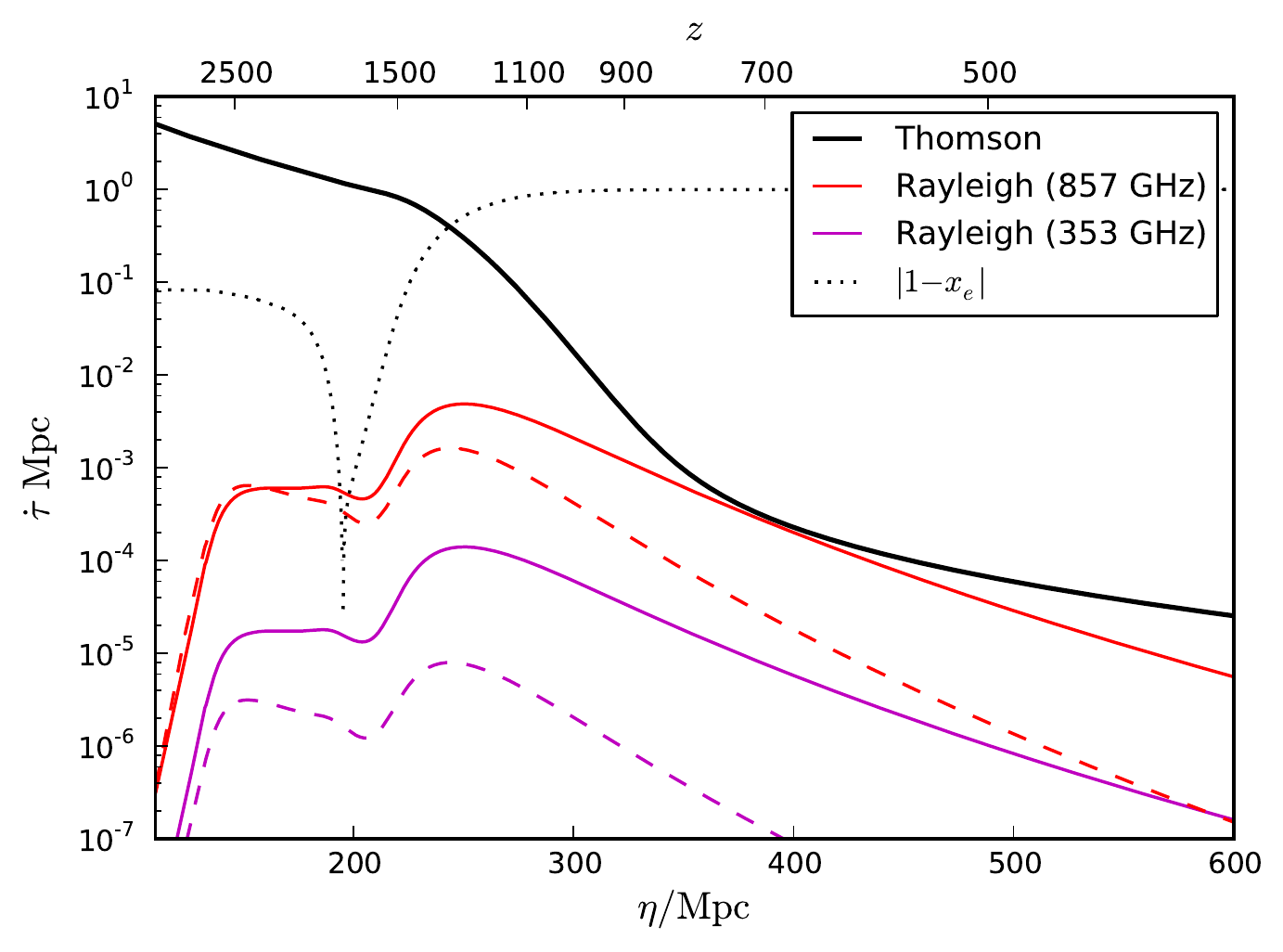}
\includegraphics[width=8.3cm]{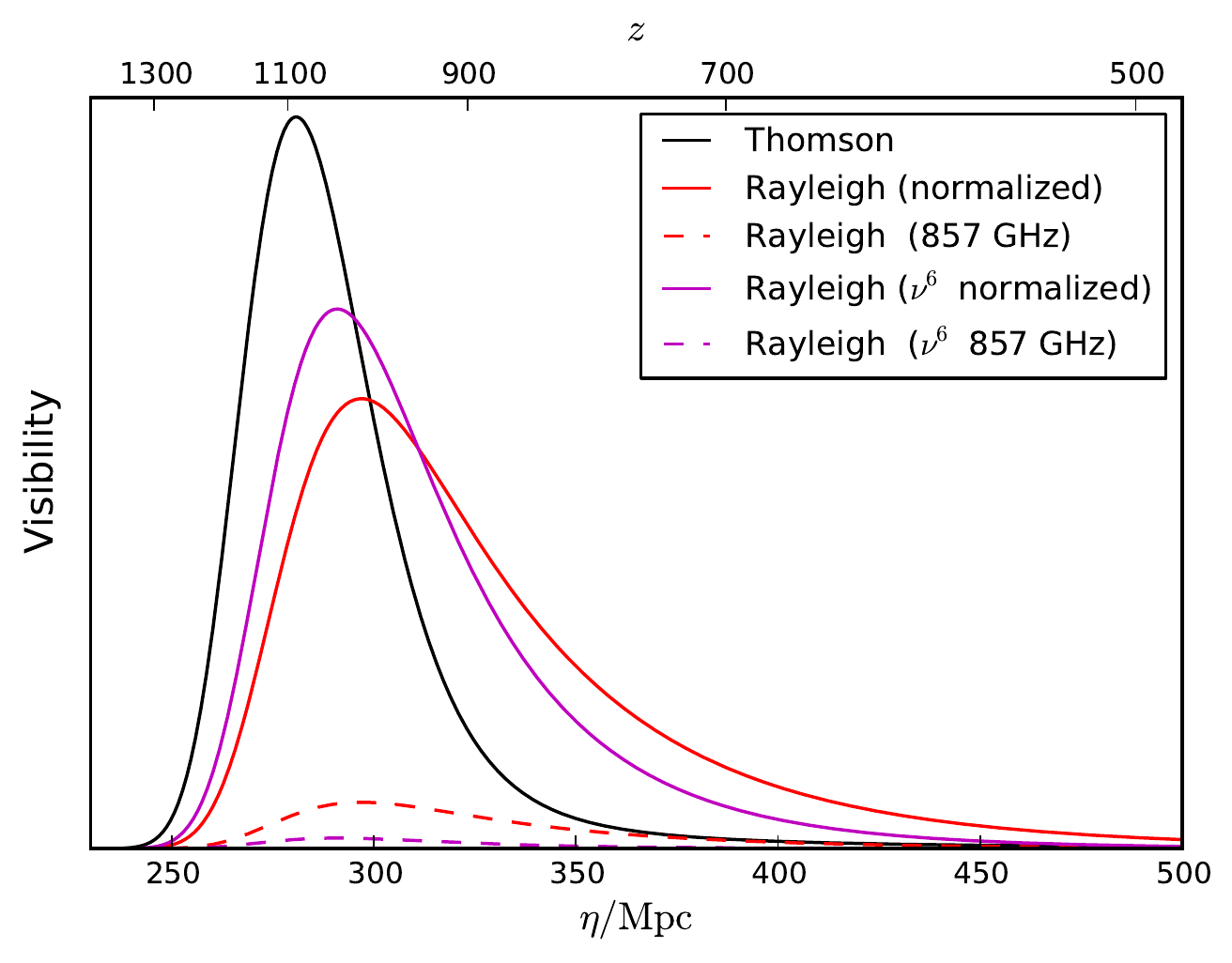}
\caption{\emph{Left:} Differential optical depth (comoving opacity $\dot{\tau}=\Gamma/(1+z)$) for Rayleigh and Thomson scattering of photons as a function of conformal time $\eta$, with Rayleigh terms scaling $\propto \nu^4$ (solid) and $\propto \nu^6$ (dashed) shown separately for a couple of observed frequencies. Before the main recombination event the Rayleigh scattering is from neutral helium, which is highly subdominant to Thomson scattering. The Rayleigh opacity decays rapidly with redshift due to the $\nu^4 \propto (1+z)^4$ redshifting and the $(1+z)^3$ dilution of the gas density with the expansion.
\emph{Right:} The corresponding visibility functions. The solid lines have been normalized; the dashed line shows the relative amplitude of the Rayleigh scattering for $857\GHz$ (which is about the upper limit of observationally relevant frequencies). Here the visibility is defined as $\dot{\tau} e^{-\tau_{\rm tot}}$.
\label{taudot}
}
\end{center}
\end{figure*}

The effect on the temperature anisotropies has been previously calculated by Refs.~\cite{Takahara91,Yu:2001gw} and shown to give several percent effect on the power spectra at $550 \GHz$ and above. The Rayleigh scattering effect becomes stronger at higher frequencies, but of course the blackbody spectrum is also falling rapidly, so there are not many observable photons at very high energies. Fig.~\ref{rms_spectra} shows the CMB brightness intensity as a function of frequency for the primary and Rayleigh anisotropies, which shows that the Rayleigh signal is most likely to be observable over a range the range of frequencies $200\Ghz \alt \nu \alt 800\Ghz$ in the absence of foregrounds. This range is spanned by \planck\ where the signal may be detectable (subject to calibration and foreground issues), and could also be measured at much higher sensitivity by a next-generation space CMB mission.

Rayleigh scattering is important for a couple of reasons: firstly, it is always present, so must be modelled consistently in any analysis using high-frequency channels for CMB or foreground separation analysis; secondly it may be able provide new information about the early universe, potentially tightly constraining the expansion rate and ionization history around recombination, and also probing additional primordial perturbation modes.
 In this paper I extend the previous calculation of Ref.~\cite{Yu:2001gw} to model the Rayleigh temperature signal in more detail, provide a new calculation of the polarization signal, and discuss future detectability and measurement.

 The outline of this paper is as follows: Sec.~\ref{scatter} reviews the details of Rayleigh scattering and the scattering sources for the CMB; in Secs.~\ref{temp} and~\ref{pol} give numerical results for the temperature and polarization respectively, and Sec.~\ref{approx} describes the approximate form and contributions to the auto and cross power spectra. Sec.~\ref{detect} discusses the ideal detectability of the signal with both current data and possible configurations for a future space mission, and Sec.~\ref{info} then studies whether an accurate measurement of the Rayleigh signal could be used to extract more information about the primordial perturbations. I assume a linearly-perturbed standard $\Lambda$CDM cosmology throughout, and that with suitable sky cuts and observations at many frequencies foregrounds can be subtracted accurately. The intricate work required to assess likely realistic levels of foreground residuals and implementation of foreground separation technology with non-blackbody CMB spectra is deferred to the future; in any case knowledge of the expected foregrounds at the required scales and level of detail is currently rather limited, so making any clear prediction would be difficult at this stage.

\section{Rayleigh scattering}
\label{scatter}

The non-relativistic Rayleigh scattering of photons with frequency $\nu$ from hydrogen in the ground state has cross section given by~\cite{HeeWon05}
\begin{multline}
%\sigma(\nu) = \left(\frac{\nu}{\nu_l}\right)^4\left[\frac{81}{64}  + \frac{957}{256}\left(\frac{\nu}{\nu_l}\right)^2 +
% \frac{ 1 626 820 991}{
%84 934 656}\left(\frac{\nu}{\nu_l}\right)^6 \dots  \right]\sigma_T
\sigma_R(\nu) = \biggl[\left(\frac{\nu}{\nueff}\right)^4 +
%\biggl[1  +
\frac{638}{243}\left(\frac{\nu}{\nueff}\right)^6
 \\+
 \frac{1299667}{236196}
 %Corrected\frac{ 1626820991}{136048896 }
 \left(\frac{\nu}{\nueff}\right)^8 +\dots  \biggr]\sigma_T
\label{series}
\end{multline}
for $\nu \ll \nueff$, where $\sigma_T$ is the Thomson scattering cross section and $\nueff \equiv \sqrt{8/9} c R_A\approx 3.1\times 10^6 \GHz$ (where $R_A$ is the Rydberg constant, corresponding to the Lyman limit frequency).
The result is derived from the Kramers--Heisenberg formula not including the intrinsic quantum mechanical line width of the excited levels, and hence should not be used for scattering close to or above the Lyman-$\alpha$ frequency where resonant scattering becomes relevant (for $z\alt 1500$ this requires observed frequencies $\nu \alt 1600\GHz$).
The corrections to the long-wavelength $\nu^4$ scaling become non-negligible around recombination, giving a greater than $10\%$ contribution from the $\nu^6$ term at observed frequencies $\nu \agt 500\GHz$ (see Fig.~\ref{taudot}).

Rayleigh scattering is easily included in a line of sight Boltzmann code, with the total scattering rate for photons with frequency $\nu$ in the gas rest frame given approximately by
\begin{equation}
\Gamma(\nu) = n_e \sigma_T+ \sigma_R(\nu) \left[\nH  + R_{\rm He}\nHe \right].
\end{equation}
Here $n_{\rm H}$ and $n_{\rm He}$ are the number densities of neutral hydrogen and helium, and $R_{\rm He}\approx 0.1$ is the relative strength of Rayleigh scattering on helium compared to hydrogen~\cite{Tarafdar69}.
I calculate the ionization history (and hence $n_e$, $\nH$ and $\nHe$) using the approximate recombination model of Ref.~\cite{Seager:1999km} calibrated to full multi-level atom codes~\cite{AliHaimoud:2010dx,Chluba:2010ca}.

The total scattering rate is very insensitive to the helium  modelling since its contribution to total scattering is very small before recombination (see Fig.~\ref{taudot}), and only a percent level correction once there is a significant amount of neutral hydrogen. Scattering from ionized helium and other constituents can be neglected, and here the relevant frequencies are well above the 21cm hyperfine transition energy (the 21cm signal following recombination is considered in detail in Ref.~\cite{Lewis:2007kz}).
Resonant scattering from excited states of hydrogen, and other atoms and molecules at later times, can also produce  interesting frequency-dependent signals
%Lithium ~\cite{Zaldarriaga:2001qt, - but see astro-ph/0507106 says nearly zero
~\cite{Basu:2003th,RubinoMartin:2005dm,Hernandez-Monteagudo:2006ar,Schleicher:2008ji,Sunyaev:2013aoa}. The cross sections for resonant scattering can be large, but all are suppressed by very low abundance. The frequency and angular dependence of the resonant scattering signal is very different from Rayleigh scattering  (which has smooth monotonic frequency dependence), and should not be a major source of confusion in practice. Resonant scattering is not included in the results of this paper.

I calculate numerical power spectra using a modified version of \CAMB\footnote{July 2013 version; modified code on the {\sf rayleigh} branch of the git repository (access available on request).}~\cite{Lewis:1999bs}.
The modifications to include the Rayleigh signal are straightforward, but require the evolution of a separate Boltzmann hierarchy for each frequency of interest, each with different scattering sources, visibility and line of sight integral. The effect of the additional total baryon-photon coupling can be included using a dense sampling of frequencies (or using an effective frequency-averaged cross section~\cite{Hannestad:2000fy}), but for most purposes this effect is small enough to neglect and then only the frequencies of interest need to be evolved. Since the signal is small, for numerical stability Rayleigh-difference hierarchies can be used, giving directly the auto and cross power spectra of the Rayleigh and primary signals. Since Rayleigh scattering is only important once recombination starts, the Rayleigh hierarchies are only evolved once the tight coupling approximation is turned off. Note that the frequency dependence of the Rayleigh cross section does not introduce any additional terms due to boosting from the gas rest frame at linear order because the net scattering effect is zero in the background.

Rayleigh scattering is only negligible compared to Thomson scattering when
\be
\nH\left( \frac{[1+z]\nu}{3\times 10^6 \GHz}\right)^4 \ll n_e.
\ee
As recombination happens $n_e$ drops rapidly, which increases the relative importance of Rayleigh scattering even though the frequencies of interest are significantly below $\nueff$ at the time.
For CMB observations Rayleigh scattering is potentially important for the higher end of observable frequencies, i.e. $\nu \agt 200\GHz$, even though it is only a small fractional change. Detectability is limited by noise (and foregrounds), not cosmic variance of the primary anisotropies, since the same perturbation realization is being observed at the different frequencies and the Rayleigh signal is strongly correlated to the primary CMB~\cite{HernandezMonteagudo:2004xg}.

%\begin{figure}
%\begin{center}
%\includegraphics[width=\hsize]{rayleigh_term_effect.pdf}
%\caption{Fractional error on the Rayleigh scattering cross section as a function of redshift neglecting $(\nu/\nueff)^6$ (solid)
%and $(\nu/\nueff)^8$ (dashed) terms.
%\label{seriesterms}
%}
%\end{center}
%\end{figure}

\section{Rayleigh temperature signal}
\label{temp}
Neutral hydrogen is only generated once recombination is underway, so the visibility function for Rayleigh scattering is peaked at somewhat lower redshift than the main Thomson scattering signal as shown in Fig.~\ref{taudot}. However photon frequencies redshift so that $\nu^4\propto (1+z)^4$, and densities dilute $\propto (1+z)^3$ due to expansion, so the amount of Rayleigh scattering does decay rapidly with time: the visibility is still well localized around the last scattering surface. In principle it probes slightly different perturbations to the primary signal during recombination. However the signal is highly correlated to the primary anisotropies, and since the Rayleigh signal is small the dominant detectable signal is the correlation of the Rayleigh contribution with the primary CMB, though there is also a small uncorrelated component (see Sec.~\ref{info}).    The Rayleigh scattering contribution originates from a somewhat later time that the primary visibility peak, so its contribution has acoustic oscillations shifted to slightly lower $l$. The total power differs from the primary signal by both an oscillatory structure, and also a power decrement on small scales since a given fixed $l$ is damped more.

\begin{figure}
\begin{center}
\includegraphics[width=\hsize]{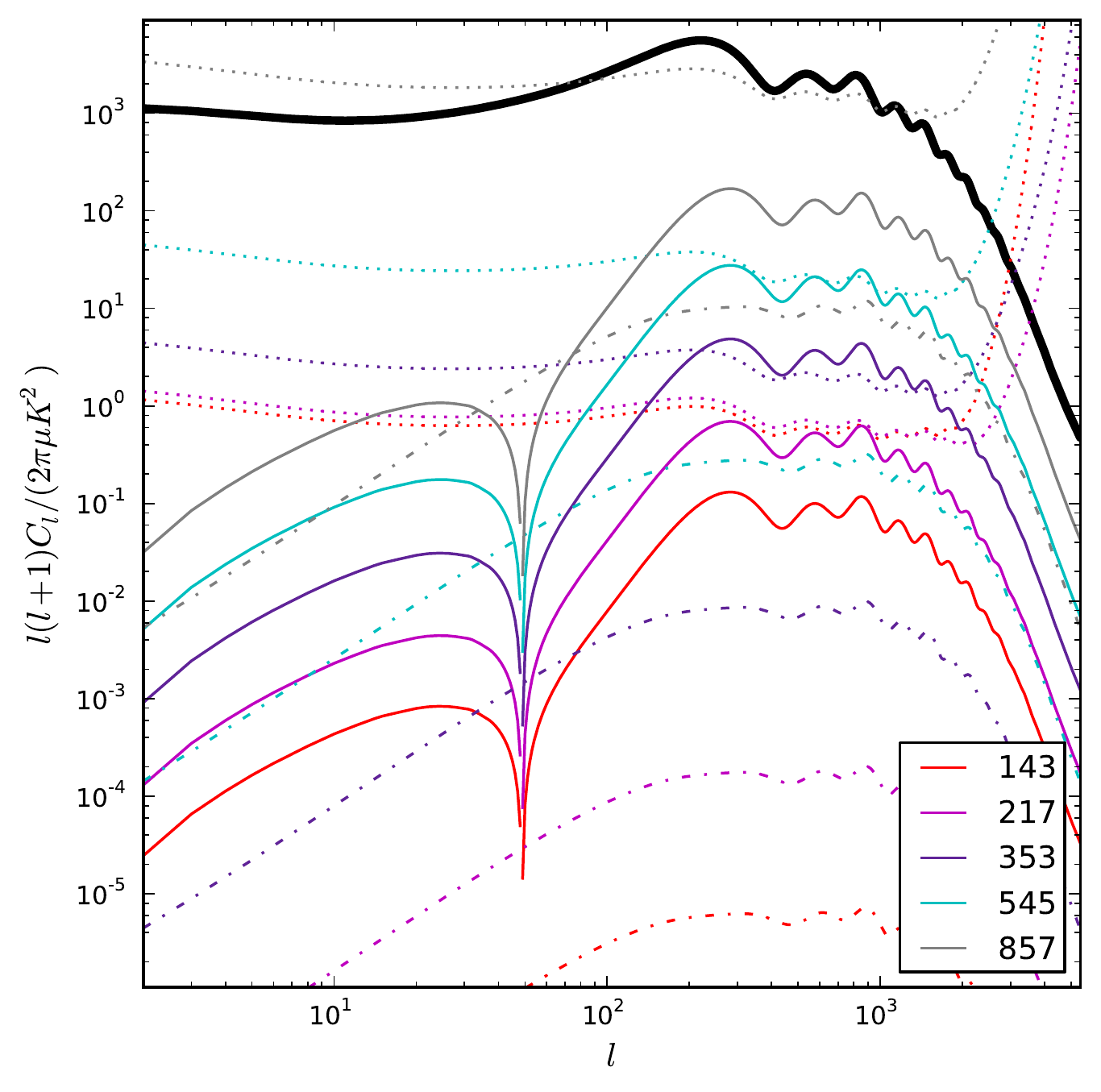}
\caption{Rayleigh contributions to the temperature power spectra at \Planck\ frequency channel notional central frequencies (in $\GHz$, colours). Solid lines are the Rayleigh-primary cross correlation (scaling approximately $\propto \nu^4$), dash-dot lines show the Rayleigh-Rayleigh power spectra (scaling $\propto \nu^8$). Dotted lines show the naive error per $\Delta l=l/10$ bin in the cross-correlation (no foregrounds).
Only the cross-correlation signal is potentially detectable by \Planck.
%, most likely at $353$GHz and $545$GHz.
\label{planckTT}}
\end{center}
\end{figure}

%For the temperature this is like multiplying the $a_{lm}$ by $1+\epsilon_\nu$  where $\epsilon_\nu$ is a frequency-dependent scaling in each channel. In the approximation in which the Rayleigh scattering contributes little to the optical depth to the visibility peak (excellent at $\nu \le 545$) the scaling is simply $\epsilon \propto \nu^4$ (which should of course be appropriately integrated over each frequency bandpass).

  It is often useful to think of the high frequency observations being the sum of a primary and Rayleigh contribution, so that the total power spectrum is a sum of the primary spectrum, twice the Rayleigh-primary correlation spectrum, and the Rayleigh-Rayleigh auto-spectrum. The cross-correlation signal can easily be isolated in principle by cross-correlating a high frequency and low-frequency map (with negligible Rayleigh contribution).
 Numerical results for the temperature auto- and cross-spectra are shown in Fig.~\ref{planckTT} for various frequencies, compared to an idealized \planck\ error model. The cross-correlation spectrum dominates the observable signal for current generation observations like \planck.

Rayleigh scattering also increases the total coupling between photons and baryons, which affects the perturbations at all frequencies, e.g. via the baryon velocity. This effect is very small, $\sim 0.04\%$ (in agreement with Refs.~\cite{Takahara91,Hannestad:2000fy,Yu:2001gw}), and can be neglected for current observations (and is anyway not frequency dependent). The slowing of baryon cooling is also negligible because the energy transfer in recoil from hydrogen is much lower than from a much lighter electron.
For further discussion and a more detailed semi-analytic discussion of the approximate form of the Rayleigh scattering temperature signal see Ref.~\cite{Yu:2001gw}.
\section{Polarization}
\label{pol}

\begin{figure}
\begin{center}
\includegraphics[width=\hsize]{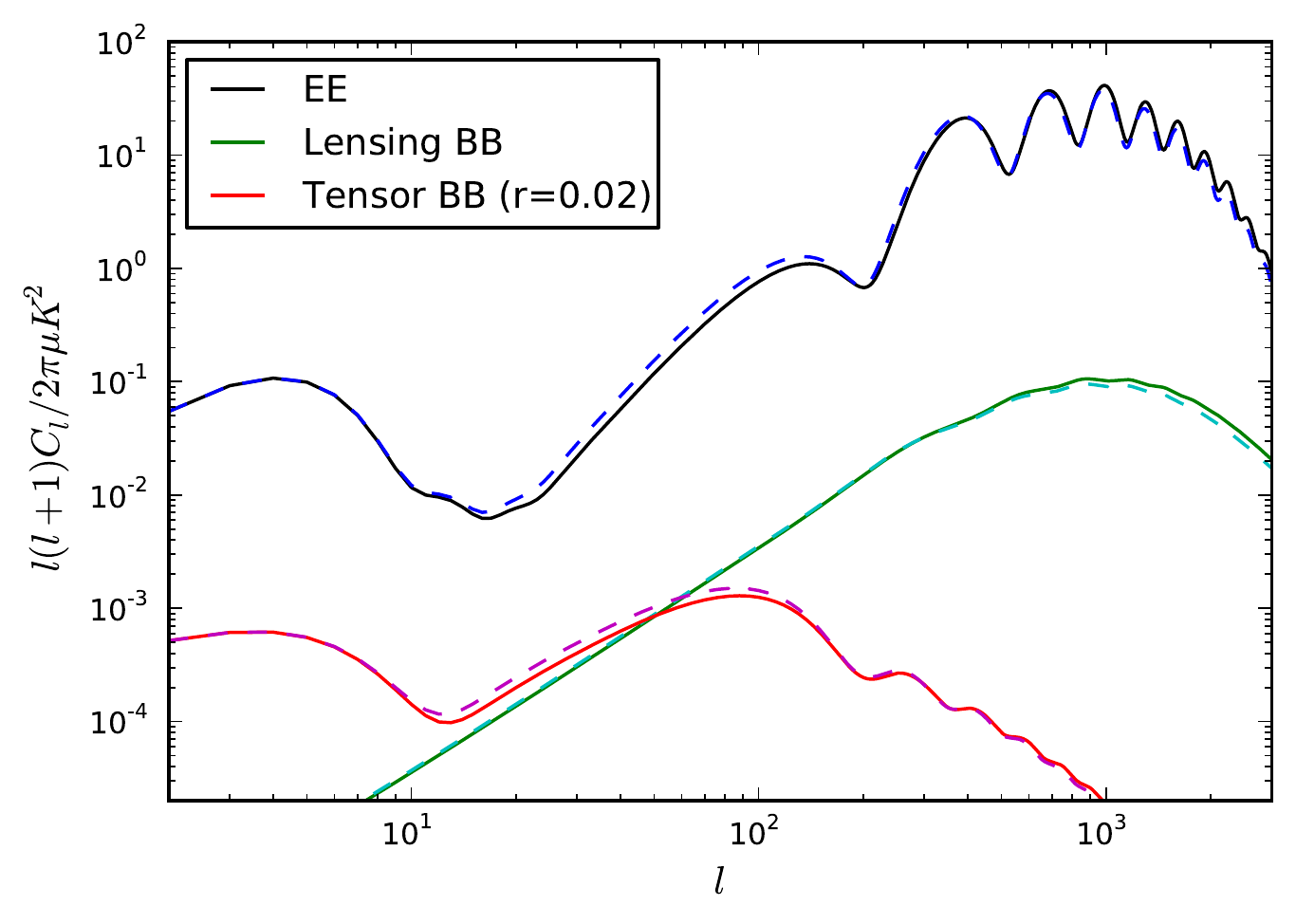}
\caption{Lensed polarization power spectra at low frequencies where Rayleigh scattering is negligible (solid) and $857\GHz$ (dashed).
The latter high frequency is chosen to see the Rayleigh contribution by eye but may not be observable in practice. The low-$l$ reionization and lensing signals are hardly changed, but the $10\alt l \alt 100$ polarization power is significantly boosted, along with a significant suppression in the damping tail at high $l$.
\label{polspectra}}
\end{center}
\end{figure}

Rayleigh scattering is also polarized: in the classical limit the scattering from the induced dipole has the same
$d\sigma_R \propto |\vepsilon_1\cdot \vepsilon_2|^2$  structure as Thomson scattering, where $\vepsilon_i$ are the polarization vectors. This should be a good approximation to energies much larger than those of relevance for the CMB since spin-flip scattering events are highly suppressed even at high energies~\cite{Safari12}. Hence the
Rayleigh polarization can be handled in a Boltzmann code in exactly the same way as Thomson scattering, e.g. following Refs.~\cite{Ma:1995ey,Hu:1998mn,Challinor:2000as}.

\begin{figure*}
\begin{center}
\includegraphics[width=\hsize]{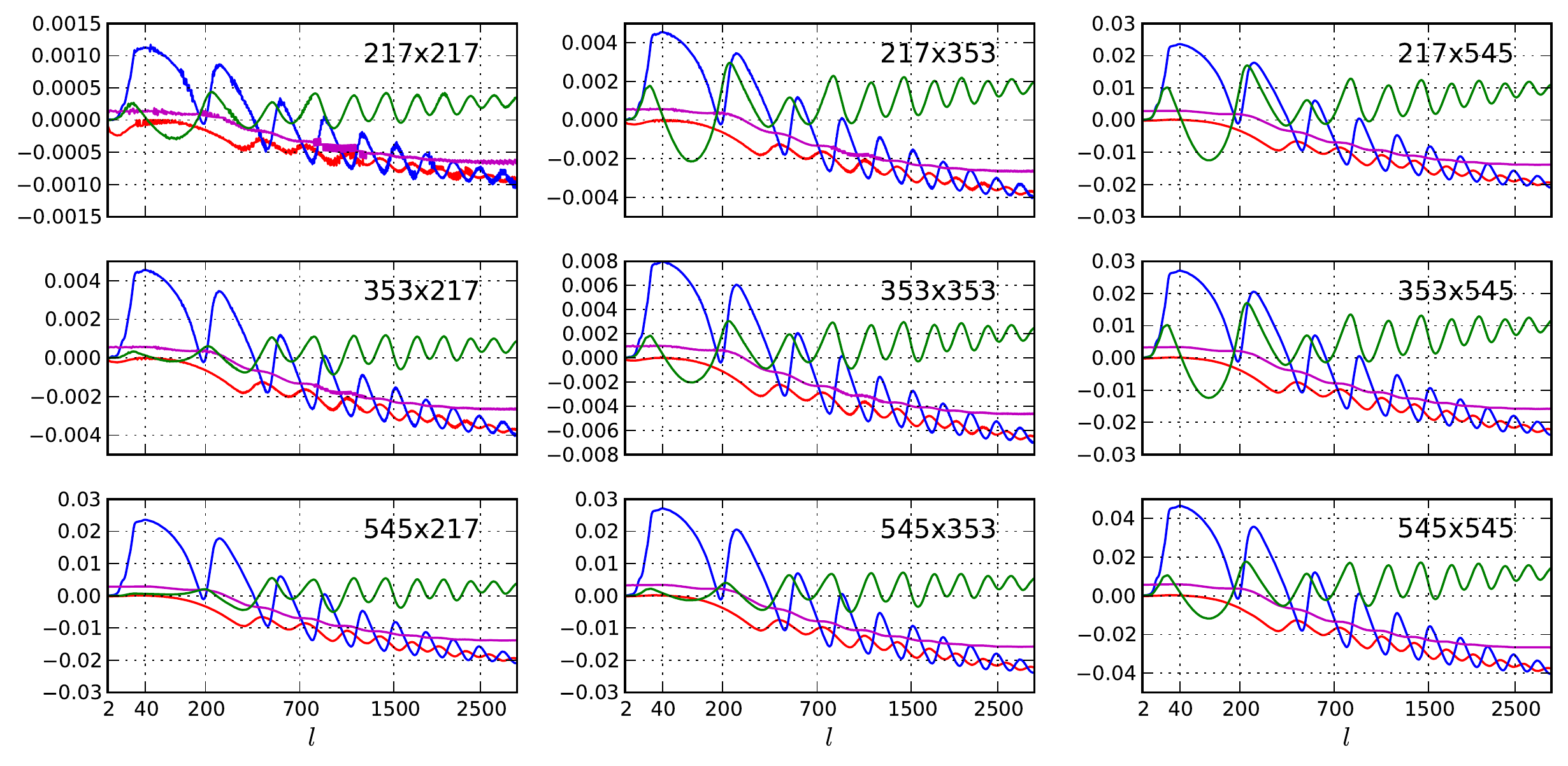}
\caption{Fractional difference between the lensed scalar CMB power spectra $C_l^{X_iY_j}$ for observed frequencies $217$, $353$ and $545$ $\Ghz$, compared to the primary (low-frequency) power spectra. Each plot shows the fractional difference $\Delta C_l/C_l$ for temperature (red), $E$-polarization (blue) and $B$-polarization (magenta), and $\Delta C_l^{TE}/\sqrt{C_l^{EE}C_l^{TT}}$ for the $T$-$E$ cross-correlation spectra (green). Each plot is a different pair of frequencies, and the results above and below the diagonal are the same except for the $C_l^{T_iE_j}$ correlation (green) which is not symmetric. Note that a small fractional difference does not necessarily mean that the signal is unobservable, since detectability is only limited by noise (and foregrounds); conversely a relatively large fractional difference in the polarization is not observable unless the noise is low enough.
\label{fracDiff}}
\end{center}
\end{figure*}

Since the Rayleigh visibility peaks at later times, the horizon size there is larger, and the large-scale polarization signal in $E$-modes from Rayleigh scattering of the quadrupole has more power on large scales, giving a frequency-dependent boost to the power beyond the reionization bump; see Fig.~\ref{polspectra}. There is also a suppression of power on small scales for the same reason as in the temperature spectrum. The large-scale bump in the spectrum is due to Thomson scattering at reionization where Rayleigh scattering is negligible, and hence remains essentially unchanged. Corresponding fractional differences to the power spectra are shown in Fig.~\ref{fracDiff}. For the polarization there are differences at the several percent level on both large and small scales.

A quadrupole induced by gravitational waves entering the horizon at recombination would also Rayleigh scatter, giving a similar Rayleigh contribution to the BB tensor-mode power spectrum. In contrast the $B$ modes produced by lensing of $E$ modes originate from polarization at recombination from a wide range of scales, where the Rayleigh signal has varying sign. The Rayleigh contribution to the lensed BB spectrum therefore partly averages out giving a significantly smaller Rayleigh contribution to the lensing BB power spectrum on large scales as shown in Fig.~\ref{polspectra}.

\section{Power spectra }
\label{approx}

The in-principle direct observables are the angular power spectra between all the fields and frequencies:
\begin{equation}
C_l^{X^i Y^j} = \la X^{i*}_{lm} Y^j_{lm}\ra ,
\end{equation}
where $X$ is T, E, or B, and $i$ labels the frequency.  At very high frequencies $\nu \agt 800\GHz$ where the Rayleigh scattering becomes a substantial effect there are very few CMB photons and very high foregrounds, and the signal is unlikely to be observable in practice. At lower frequencies the Rayleigh scattering is a small effect and can accurately be modelled perturbatively as
\begin{equation}
X_{lm}^i \approx X_{lm} + \left( \frac{\nu_i}{\nu_0}\right)^4 \Delta X_{4,lm} + \left( \frac{\nu_i}{\nu_0}\right)^6 \Delta X_{6,lm},
\label{pertX}
\end{equation}
where $X_{lm}$ is the primary (low frequency) signal, $\nu_0$ is some reference frequency, and $\Delta X_i$ are the contributions due to Rayleigh scattering at frequency $\nu_0$ with the corresponding frequency scaling. This approximation will break down when higher terms in the Rayleigh scattering cross section become relevant, and also when the Rayleigh optical depth becomes significant, but is a good approximation over most of the frequency range of interest. With this approximation the power spectra are
\begin{multline}
C_l^{X^i Y^j} \approx  C_l^{XY} + \left( \frac{1}{\nu_0}\right)^4 \left[ \nu_j^4 C_l^{X \Delta Y_4^j} + \nu_i^4 C_l^{\Delta X_4^i Y}\right] \\
   \qquad\qquad + \left( \frac{1}{\nu_0}\right)^6 \left[ \nu_j^6 C_l^{X \Delta Y_6^j} + \nu_i^6 C_l^{\Delta X_6^i Y}\right]
   \\+ \left( \frac{\nu_i\nu_j}{\nu_0^2}\right)^4 C_l^{\Delta X_4^i \Delta Y_4^j}
   +\dots
\end{multline}
The leading Rayleigh term fits the difference spectra shown in Fig~\ref{fracDiff} rather well.
The last term is not required at \Planck\ sensitivity where cross-correlations dominate the observable signal, but would be detectable with future missions (see Sec.~\ref{info}), and is quantitatively more important than the $\clo(\nu^8)$ cross-correlation scattering contribution.
The observed spectra are of course lensed, and the lensed spectra can be calculated for all the cross frequencies using standard techniques~\cite{Lewis:2006fu}.
%Note that the leading $\clo(\nu^4)\times \clo(\nu^4)$ Rayleigh auto spectrum is typically significantly larger than the $\clo(\nu^8)$ scattering cross-spectrum, because the latter visibility decays very rapidly with redshift during recombination.

Since the Rayleigh signal is significant at higher frequencies, it is important to model it in cosmological and foreground analyses using those frequencies.
The signal is easily simulated exactly from the full set of cross-frequency power spectra, and also for current data approximately using Eq.~\eqref{pertX}  by writing
\begin{equation}
\vx^\nu_{lm} = \vx_{lm} + \left( \frac{\nu}{\GHz}\right)^4 \mR_{4,l}\,  \vx_{lm}
 + \left( \frac{\nu}{\GHz}\right)^6 \mR_{6,}\,  \vx_{lm} +\dots
\end{equation}
where $\vx_{lm} = (T_{lm}, E_{lm})$ is a simulation of the blackbody fields, and $\mR_i$ is a $2\times 2$ matrix that can be computed from the (unlensed) cross power spectra such that  $\vx^\nu_{lm}$ has the required covariance to leading order in the cross-correlation Rayleigh effect. The separate terms can also be individually lensed and then combined with the appropriate weighting.
%\begin{equation}
%\mR = \frac{\nueff^4}{C^{TT}_l C^{EE}_l -(C^{TE}_l)^2}
%\Begm
%C^{T\Delta T_l C^{EE}_l - C^{TE}_l C_l^{T\Delta E} &
%C^{T\Delta T_l C^{T\Delta E}_l - C^{TE}_l C_l^{T\Delta E} &
%\enm
%\end{equation}
%For speed we may in the first instance apply this to the lensed maps to a good approximation, rather than simulating and then lensing.

\section{Detectability}
\label{detect}

\begin{figure*}
\begin{center}
\includegraphics[width=6.1cm]{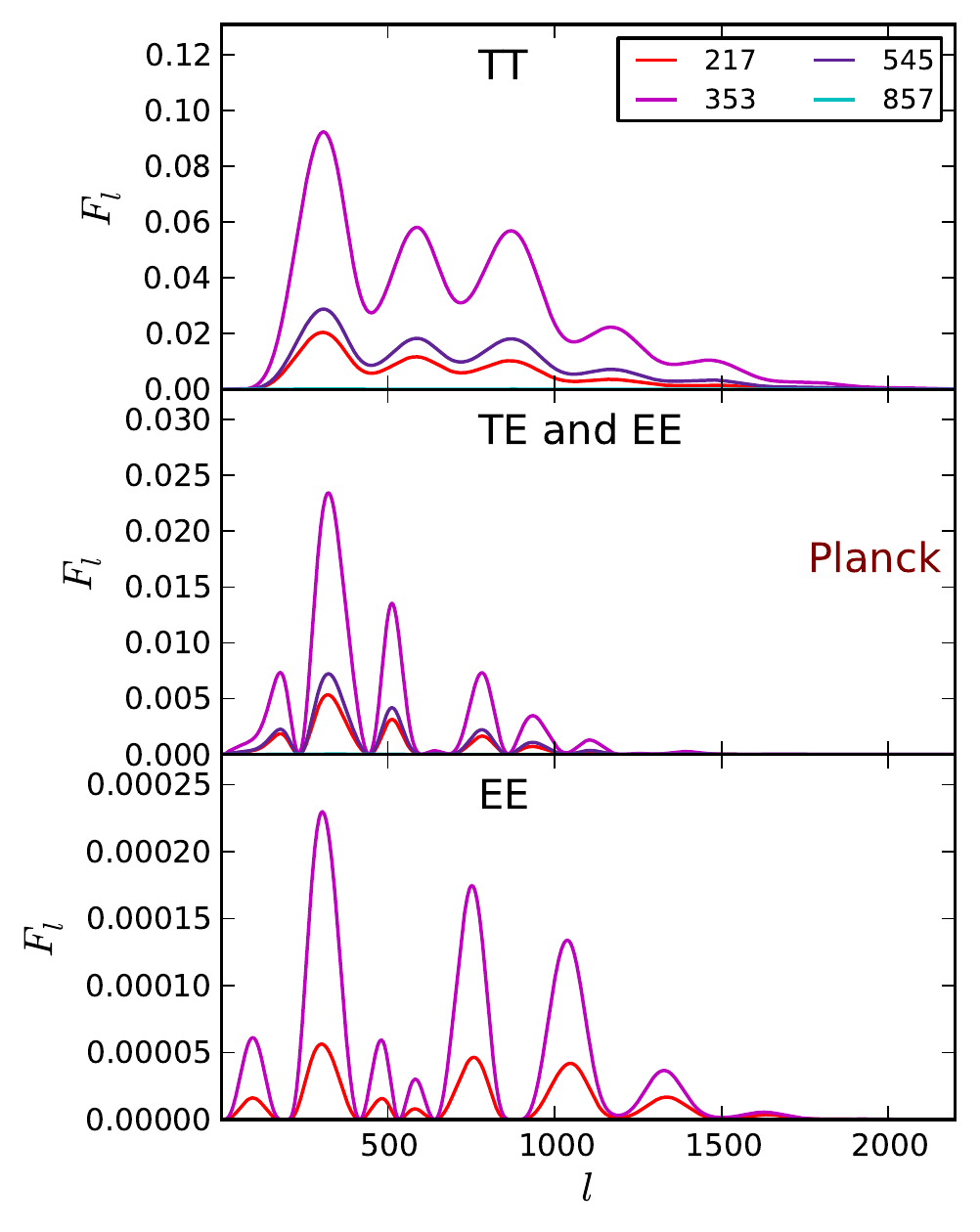}
\includegraphics[width=5.9cm]{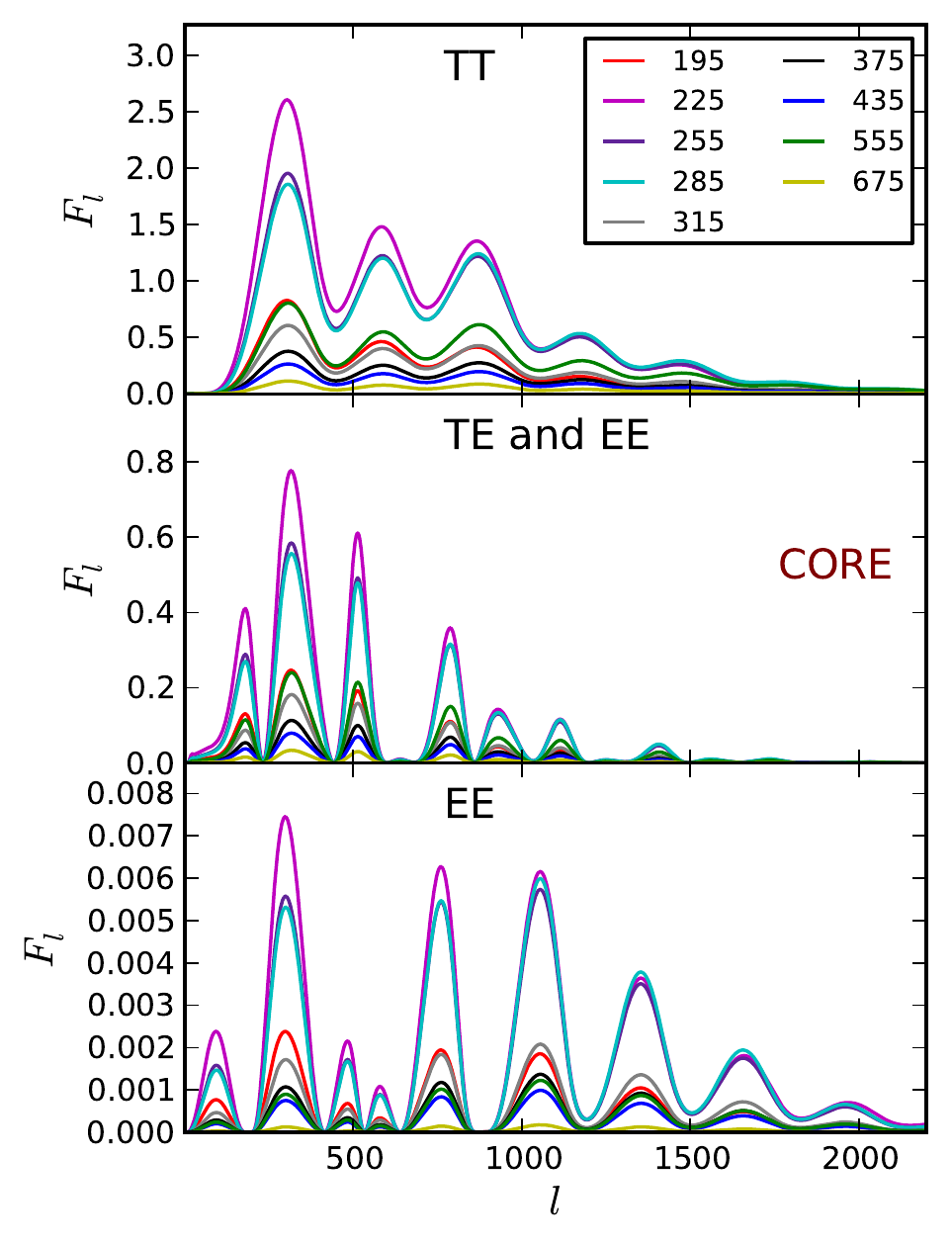}
\includegraphics[width=5.7cm]{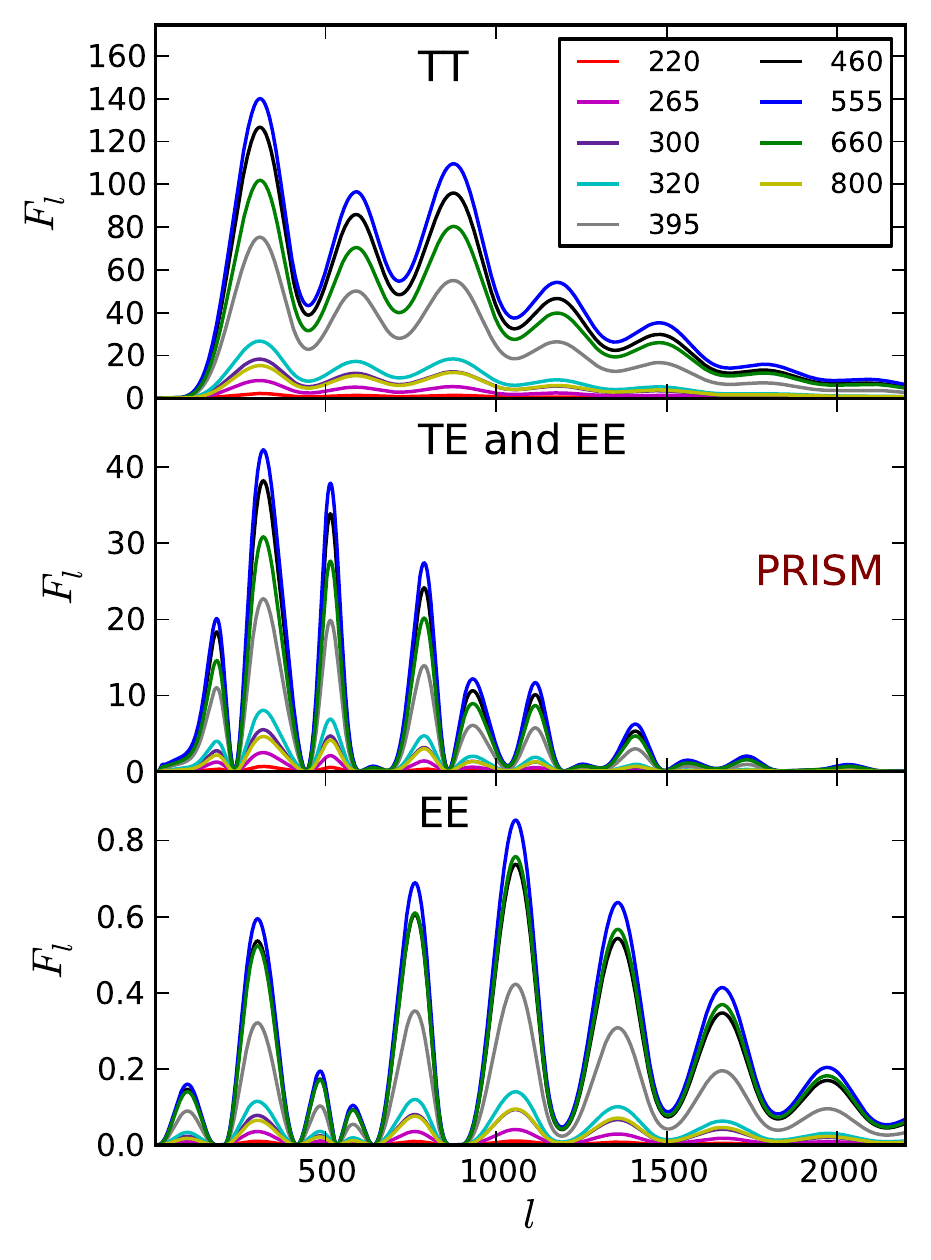}
\caption{Approximate Fisher signal to noise per $l$ for the Rayleigh-primary cross-spectra in the most sensitive high frequency channels of \Planck, and some individual broad-band high-frequency channels for a couple of configurations for proposed future space missions (\CORE\ and \PRISM). Noise is assumed to be white on small scales with some flattening at low $l$, and $\fsky=0.6$; no foreground residuals are included.\label{FisherPlot}}
\end{center}
\end{figure*}

At high frequencies measurement of the Rayleigh signal is limited by the rapid fall in the blackbody spectrum (so fewer CMB photons and hence larger relative noise), and foregrounds including dust, SZ, molecular lines and the cosmic infrared background (CIB), though the latter is expected to be only weakly polarized. A detailed treatment of likelihood uncertainties from foreground modelling is beyond the scope of this paper, but there are several reasons why they may not be a major problem for detecting the Rayleigh signal if many high sensitivity channels are included at $100\GHz \alt \nu \alt 800\GHz$:
 \begin{itemize}
 \item The shape of the Rayleigh scattering spectra can be computed accurately (for fixed cosmological parameters).
  \item   The Rayleigh signal spans the full range of scales, so for example CIB and SZ should only be a small contaminant at lower $l$, and the dust spectrum falls to higher $l$.
 \item The dense frequency coverage and low noise levels required to clean foregrounds for detecting primordial $B$ modes
     %and look for spectral distortions
     should also be adequate to clean the higher $l$ foregrounds to within small residuals.
 \item Unlike the foregrounds, the Rayleigh signal is strongly correlated to the primordial CMB, so a cross-spectrum between a low frequency map (with low dust, CIB and noise) and a cleaned high-frequency map (with strong Rayleigh signal) is expected to have small foreground contamination.
 \item Residual foregrounds should have a power spectrum shape in $l$ that looks very different from the predicted Rayleigh contribution and will mainly serve to increase the effective noise. In particular the Rayleigh signal is oscillatory, with amplitude nearly tracking the primary power spectrum on small scales, but with slightly shifted acoustic scale due to the larger sound horizon size for Rayleigh scattering~\cite{Yu:2001gw}.
 \end{itemize}
Conversely it will be important to self-consistently model the Rayleigh contribution when doing foreground separation and data analysis.
For a more detailed discussion of foreground modelling for future CMB missions see e.g. Ref.~\cite{Dunkley:2008am,Kogut:2011xw,Andre:2013afa}.
The extent to which the residual foregrounds dominate the noise budget depends on the density of the frequency sampling and the correlation between frequencies of the various components; for the relatively sparse sampling of \planck, where for example the CIB has a significant uncorrelated component between channels~\cite{Ade:2011ap}, realistic errors are likely to be substantially larger than those estimated from instrumental noise alone. At large scales where the noise levels are very low (so the difference of maps at a low and high frequency would ideally isolate the Rayleigh signal nearly perfectly), details of non-whiteness of the noise and systematic residuals may also be important to determine the actual level of sensitivity to Rayleigh differences on large scales even if foregrounds can be accurately removed.

Here I simply give ballpark sensitivity numbers assuming isotropic noise and negligible foreground residuals for some simple cases. Since the analysis is only approximate, for simplicity I use the approximate model of Sec.~\ref{approx} with leading $\nu^4$ scattering terms, and assume delta-function frequency bandpasses at band-central frequencies. For Gaussian CMB fluctuations the ideal likelihood function is straightforward. For example consider the simple case of having one low frequency channel (with negligible Rayleigh signal) and one high frequency channel, with noise $N^0_l$ and $N^\nu_l$ respectively. The covariance of the measured temperatures $\vT=(T_0, T_\nu)$ when the noise dominates the Rayleigh-Rayleigh auto power spectrum is then approximately
\be
\left\la \vT_{lm} \vT_{lm}^\dagger \right\ra\approx
\begm C_l + N^0_l & C_l + C^{TR_\nu}_l \\ C_l + C^{TR_\nu}_l & C_l +2C^{TR_\nu}_l +
%C^{R_\nu R_\nu}_l+
N^\nu_l\enm,
\ee
where $C_l$ is the primary power spectrum and $C^{TR_\nu}_l$ the (small) Rayleigh-primary cross-correlation. The Fisher matrix for the fractional measured amplitude of $C^{TR_\nu}_l$ at a given multipole assuming it is small is then
\be
\sigma^{-2}_l \approx \frac{(2l+1)\fsky\left[C_l (N^\nu_l+N^0_l) + N^0_l(N^\nu_l+2N^0_l)\right] }{ \left[C_l (N^\nu_l+N^0_l) + N^0_l N^\nu_l\right]^2}.
\ee
As expected this blows up (i.e. definite detection) as $N\rightarrow 0$ since the primary cosmic variance fluctuations are the same in both maps.
This result can serve as a guide as to whether Rayleigh scattering is important in a given high frequency map, for example see Fig.~\ref{planckTT}. The same form holds for the $EE$ polarization power spectrum, though for higher noise levels the sensitivity for polarization is likely to be dominated by the $TE$ correlation.
More general cases can be considered using the full Fisher matrix
\be
F_{ij,l} \approx \frac{(2l+1)\fsky}{2}\Tr \left[ \mC_l^{-1} \left(\frac{\partial}{\partial \theta_i} \mC_l\right)  \mC_l^{-1} \left(\frac{\partial}{\partial \theta_j} \mC_l\right) \right]
\ee
for parameters $\theta_{i}$, $\theta_{j}$ where $\mC_l$ is the full multi-frequency temperature and polarization covariance matrix (a $2N_{\rm freq}\times 2N_{\rm freq}$ matrix in the case of $T$ and $E$ observations). Since foregrounds are being neglected I shall restrict to considering the detectability of the primary-Rayleigh cross-spectrum for various high-frequency channels individually rather than super-optimistically considering joint constraints.

I assume $\fsky=0.6$ and white noise at $l\ge 1000$ with a multiplicative $(1000/l)^{0.4}$ flattening at $l<1000$ to crudely avoid massively overweighting very low $l$ (qualitatively consistent with the \planck\ temperature noise at low $l$).
Fig.~\ref{FisherPlot} shows the corresponding naive signal to noise in the Rayleigh signal for high frequency channels in \Planck\ (full mission) and a couple of straw man configurations for future space missions: \CORE~\cite{Bouchet:2011ck} (which has many more detectors at $\nu<300\Ghz$ than at high frequencies, and hence is relatively insensitive to the Rayleigh signal), and the more ambitious \PRISM~\cite{Andre:2013afa} (which has many high frequency channels with hundreds of detectors). All are sensitive enough to detect the temperature signal at several sigma. \PRISM\ would measure the temperature signal in detail with the power at each $l$ being measured with small fractional error, and an overall determination of the amplitude of the Rayleigh signal to $\sim 0.3\%$. The Rayleigh-temperature cross primary-polarization combination dominates the detectability of polarization cross-spectra, and is easily detectable in future missions. A \CORE-like configuration would be marginally able to detect the Rayleigh-primary $EE$ cross-spectrum at a bit under $2\sigma$ per channel for $200\GHz<\nu<300\GHz$, and PRISM has the sensitivity to detect it at around $20\sigma$ in each of the $400\GHz<\nu<700\GHz$ channels.

Note that contributions from $\clo(\nu^6)$ Rayleigh scattering terms become important to model for high-sensitivity observations, with PRISM being in principle sensitive to them at the $10\sigma$--$20\sigma$ level in the temperature spectra for $450\Ghz \alt \nu \alt 800\GHz$. The $\clo(\nu^8)$ scattering corrections are an additional $\clo(1\sigma)$ correction on top of that, which is unlikely to be very important in practice but is easily included in a full analysis.

\section{Additional information?}
\label{info}

\begin{figure*}
\begin{center}
\includegraphics[width=8cm]{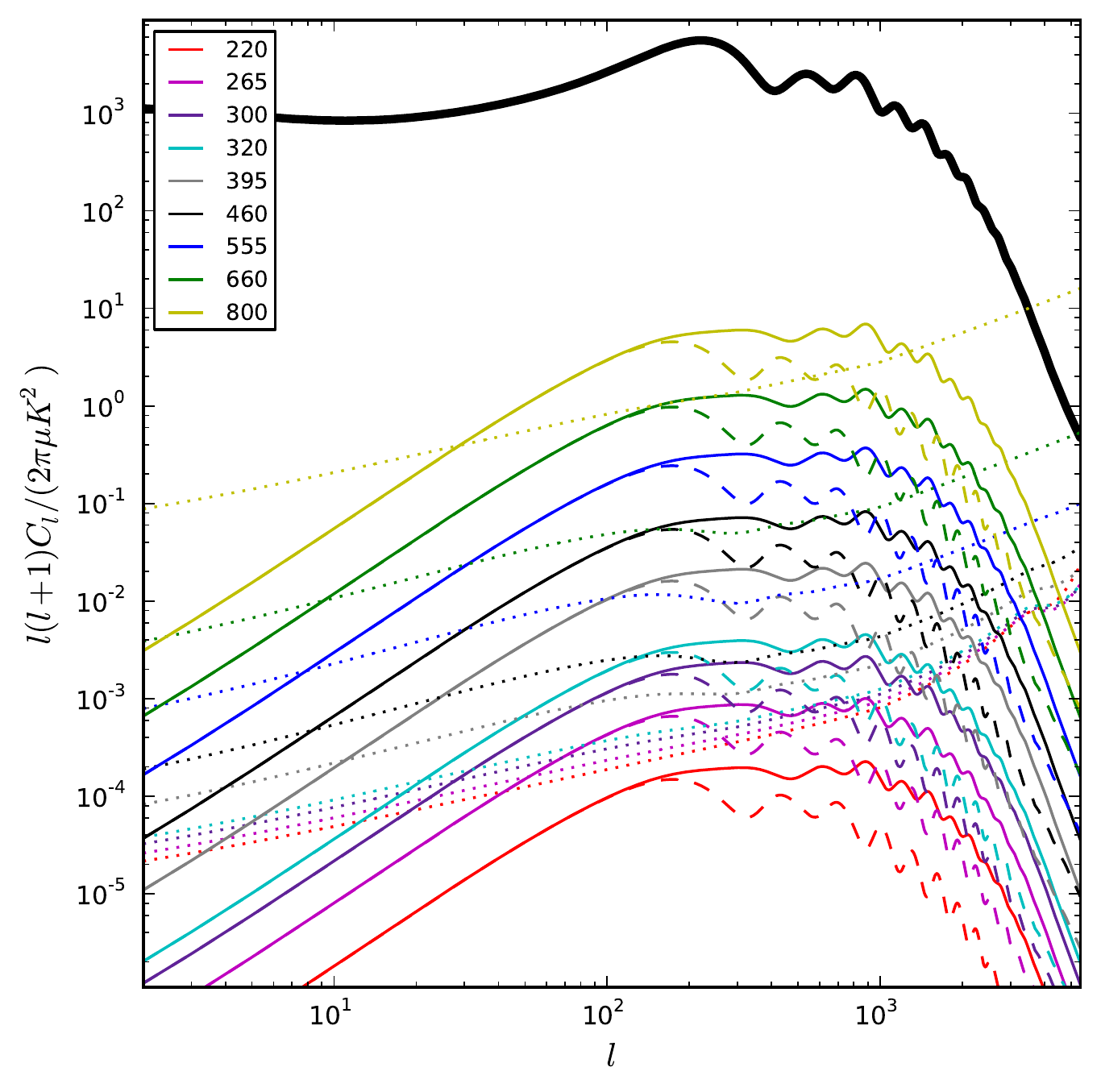}
\includegraphics[width=8cm]{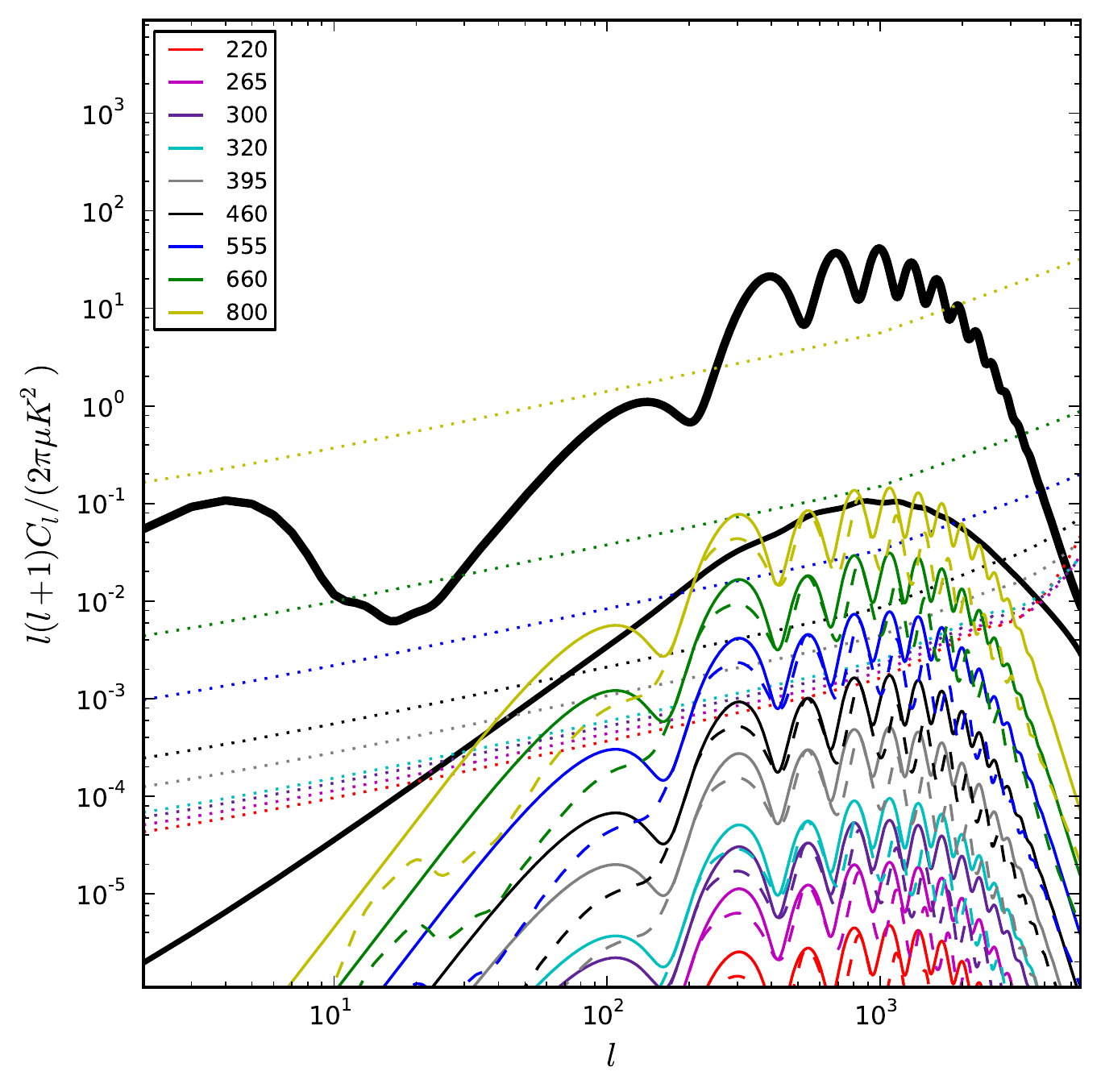}
\caption{ The Rayleigh-Rayleigh auto-spectra (solid lines), and the uncorrelated component (dashed lines) for the CMB temperature (left) and $E$-polarization (right) at various frequencies (in $\GHz$). Dotted lines show the naive zero-foreground error per $\Delta l=l/10$ bin at each frequency for a PRISM-like observation. The temperature spectrum is above the noise levels on intermediate scales, but the polarization signal would require significantly lower noise to be measured well. Thick black lines show the primary power spectra (and $C_l^{BB}$ in the case of the thinner line in the right-hand plot).
\label{PRISM_auto}}
\end{center}
%\vspace{0.5cm}
\end{figure*}

\begin{figure}
\begin{center}
\includegraphics[width=\hsize]{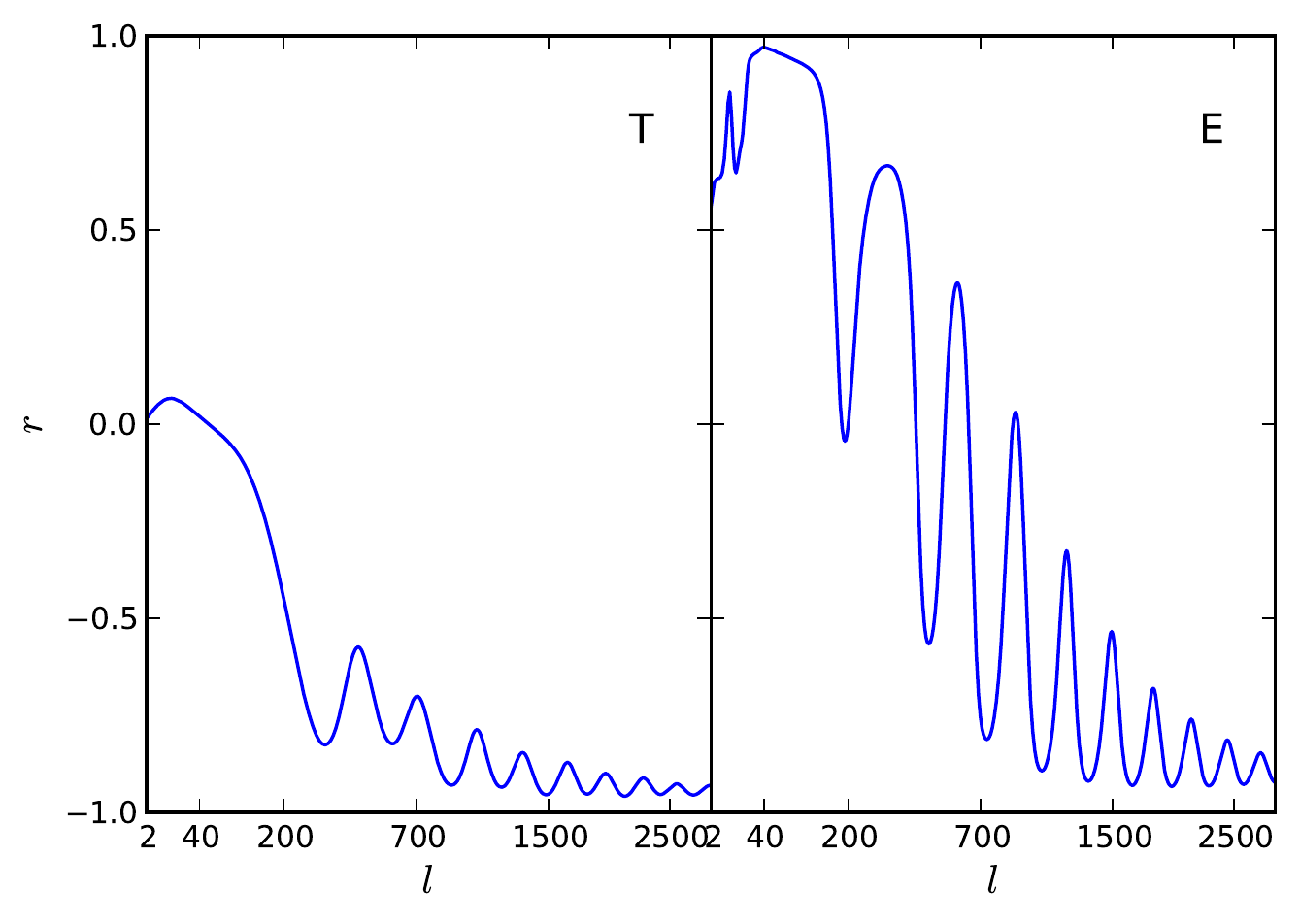}
\caption{ Correlation coefficient between the Rayleigh and primary signals (at $545\Ghz$), for temperature (left) and $E$-polarization (right).
\label{correlation}}
\end{center}
\end{figure}

On small scales the Rayleigh scattering functions mostly as an additional screen just in front of the primary last-scattering surface, leading to additional scattering and hence damping of small scale primary anisotropies. The Rayleigh signal is therefore highly (anti-)correlated to the primary signal, being largely proportional to it, with small additional contributions from anisotropies being sourced, for example by Doppler terms due to peculiar motion. The screening effect is much like the optical depth suppression from reionization, except that here the relevant horizon size is that at recombination, leaving a significantly larger range of observable super-horizon modes that are not damped in the same way.
The very large-scale $E$-mode polarization signal is also highly correlated to the primary $E$-mode signal, since it is caused by scattering from nearly the same large-scale quadrupole; see Fig.~\ref{correlation}.

On intermediate scales, or with high sensitivity, the Rayleigh scattering signal can probe different perturbation modes that cannot be isolated from the primary anisotropies, and hence contains additional information about the primordial fluctuations. The Rayleigh-primary cross-correlation signal strongly constrains the cosmological model around recombination, but does not measure new fluctuations. The uncorrelated part of the Rayleigh-Rayleigh power spectrum is what contains independent information about the perturbations. It is far too small to be measured by \Planck\, and only marginally by a \CORE-like experiment, but more ambitious future observations may be able to measure it if foregrounds can be modelled to high accuracy.

The Rayleigh anisotropies are never perfectly correlated to the primary anisotropies, so in principle each Rayleigh mode from each $\nu^n$ term in the cross section has additional information. A perfect measurement of a small $\nu^4$-scaling effect could double the number of modes that can be measured (in the same way that polarization doubles the number compared to having just temperature). In practice, if the signals are small and also highly correlated, extremely low noise levels would be required to achieve this, and in practice it is likely to be impossible, especially considering the vastly larger foregrounds that must be distinguished. Nonetheless it is worth briefly considering what extra information might be available under marginally more realistic assumptions.

Fig.~\ref{PRISM_auto} shows the Rayleigh auto-spectra for the temperature and polarization at various frequencies, and also the uncorrelated component.
Even with ambitious PRISM-like observations the $E$-polarization auto spectrum is too small to be useful. However the temperature auto spectra could be above the instrumental noise, and the Rayleigh auto-spectrum measured statistically on intermediate scales. Let's define the number of modes as being
\be
n_l \equiv (2l+1) \fsky\Tr \left[ \left( [\mC_l+\mN_l]^{-1} \mC_l\right)^2 \right],
\ee
where $\mC_l$ is the matrix of theory spectra and $\mN_l$ is the corresponding noise contribution.
This definition corresponds to (two times) the Fisher matrix for an overall power spectrum amplitude parameter at a given $l$, and is the sum of the squares of the signal-to-signal-plus-noise eigenmode eigenvalues. With zero noise the primary spectra have $n_l=(2l+1)$ and $n_l=2(2l+1)$ for temperature and temperature+polarization respectively. Including a high-frequency spectrum with the Rayleigh signal then adds additional modes if the spectrum is not noise dominated, and there is a significant uncorrelated component.

\begin{figure}
\begin{center}
\includegraphics[width=\hsize]{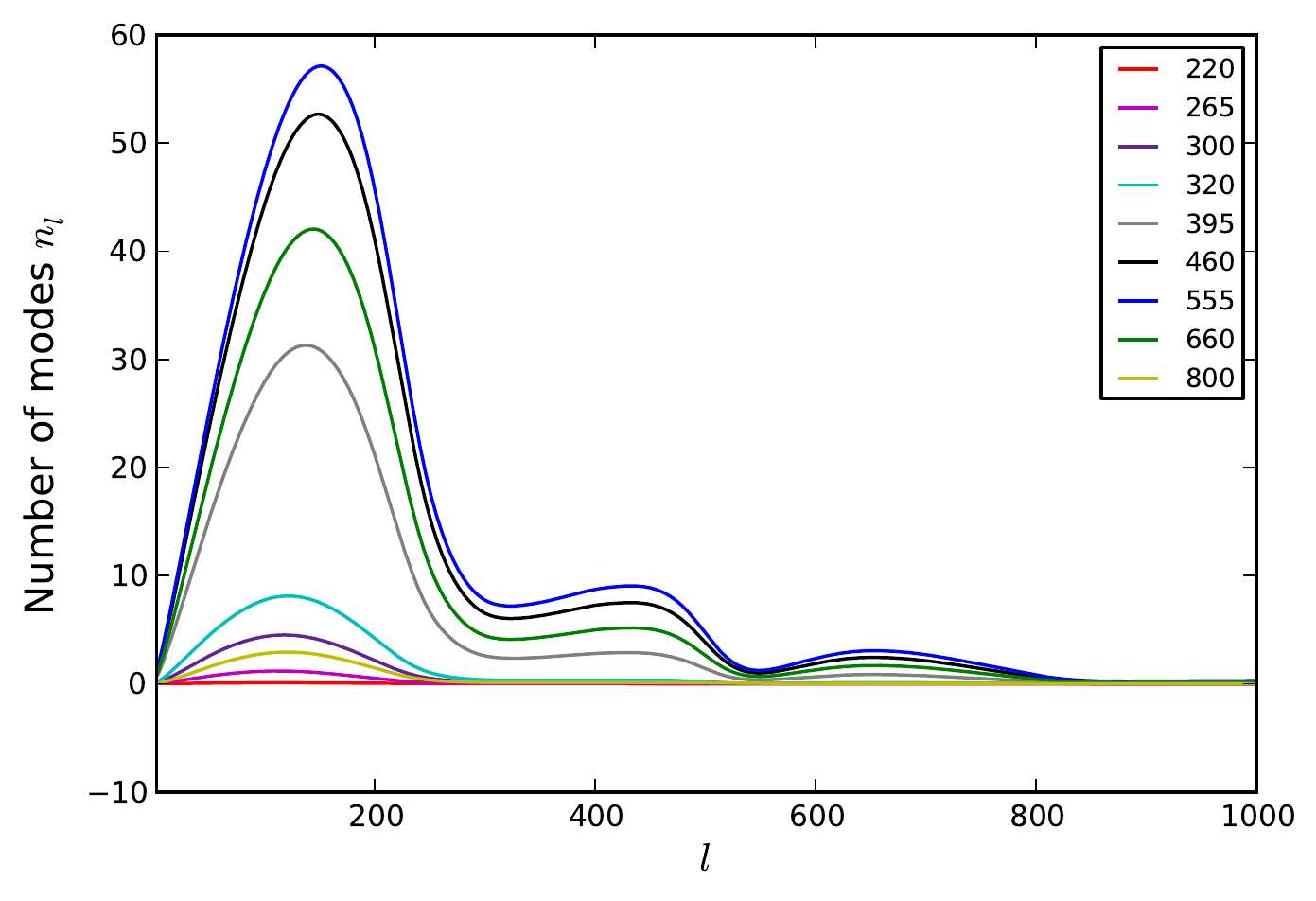}
\caption{ The number of new perturbation modes per $l$ in principle measurable by the Rayleigh scattering signal from various single broad-band high-frequency channels with \PRISM-like noise. The information is essentially all in the temperature, not the polarization, and nearly the same underlying modes are probed by each frequency.
\label{PRISM_modes}}
\end{center}
\end{figure}

With \PRISM\ sensitivity the low frequency channels have $n_l\approx 2(2l+1)$ up to high $l$ since temperature and polarization would both be measured at high signal to noise. Fig.~\ref{PRISM_modes} shows the number of additional modes as a function of $l$ when adding a high-frequency channel to probe the Rayleigh scattering signal. In total the Rayleigh measurement could probe around 10 000 new modes, mostly at $l\alt 500$. The information is on larger scales because that is where the noise is lowest, and also because the primary temperature anisotropies there come from multiple sources, which the additional Rayleigh measurement can help disentangle. For example there is no Sachs-Wolfe or Integrated Sachs-Wolfe (ISW) contribution to the Rayleigh signal because there is no Rayleigh monopole background: it is only generated by sub-horizon scattering processes, not line-of-sight redshifting effects.

Some modes contributing to the large-scale CMB also measurable by other means, for example large-scale structure and CMB lensing can probe the modes contributing to the ISW effect independently of the CMB temperature. The Rayleigh signal is however a much more direct probe of last scattering than doing model-dependent ISW inference, and most of the additional modes in this case are localized at last scattering, not ISW. In particular the large-scale Rayleigh temperature signal depends directly on the Doppler scattering terms, which are only a subdominant component of the primary anisotropies. This information is complementary to the large-scale primary polarization because the $\sim \vnhat\cdot \vv_b$ terms that source the Rayleigh Doppler signal come from modes aligned with line of sight $\vnhat$, whereas the $E$ polarization comes from the quadrupole caused by infall in directions transverse to the line of sight, so the modes being probed are nearly independent.

The number of additional modes is small compared to the total number available in the primary anisotropies ($\clo(\lmax^2)$), even with ambitious observations under simplistic assumptions. However they could in principle be useful since cosmic variance is very limiting on large scales, and there are a number of `anomalies' claimed in the large-scale temperature distribution. If measurement of the Rayleigh signal at $l\alt 200$ were achievable it would be one way to get slightly better statistics and help to separate different possible physical models, in addition to the information available in polarization. Orders of magnitude higher sensitivity than \PRISM\ would be required to get substantial additional information on smaller scales or from the polarization. However since there is potentially useful information on fairly large scales, it may also be possible to exploit degree-resolution observations with a Fourier Transform Spectrometer (FTS), as proposed by \PIXIE~\cite{Kogut:2011xw} and then \PRISM.

\section{Conclusions}

Rayleigh scattering produces an interesting detectable signal in the CMB temperature and polarization at frequencies $\agt 200\Ghz$. In summary Rayleigh scattering:
\begin{itemize}
\item Can easily be modelled accurately using a full set of cross-frequency power spectra calculated in linear theory from a Boltzmann code
\item Produces a few-percent damping of the temperature power at high frequencies that may be detectable with current observations~\cite{Yu:2001gw}.
\item Increases the total coupling of baryons and photons, leading to a very small $\alt 0.04\%$ frequency-independent increase in the small-scale primary CMB power spectra and a smaller change in the matter power spectrum~\cite{Yu:2001gw,Hannestad:2000fy}.
    \\
\item Boosts the large-scale $E$-polarization at $10\alt l\alt 300$ due to the increased horizon size for the later Rayleigh scattering.
\item Damps the small-scale polarization in a similar way to the temperature anisotropies
\item Must be modelled for consistent foreground separation, including significant corrections from the $\nu^6$ term in the cross section at higher frequencies.
\item Produces temperature and polarization signals that could both easily be measured by a future space-based mission; better measurement motivates more sensitivity in the $400\GHz \alt \nu \alt 700\GHz$ frequency range.
\item Enhances the $10\alt l\alt 100$ $B$-mode polarization signal from gravity waves at higher frequencies, and has a much smaller effect on the non-linear $B$-modes from lensing.
\item Is strongly correlated to the primary signal, so a measurement of the cross-correlation may provide a robust means of measurement and allow powerful constraints on the expansion and ionization history of the universe around recombination.
\item Probes new primordial perturbation modes, though due to the strong correlation of the bulk of the Rayleigh signal this requires very high sensitivity and foreground rejection efficiency, and is likely to be of most use for probing roughly horizon-scale modes at recombination.
\end{itemize}

Future work is required to assess likely levels of residual foreground contamination for different possible
observation strategies, and hence levels of precision that may be achievable in practice. If spectral distortions
in the monopole are studied at high sensitivity over clean degree-scale patches of sky, the local distortion due to Rayleigh scattering may also be non-negligible at the Jansky level.
%Alternatively it may be possible to increase the resolution of a Fourier Transform Spectrometer to measure the large=scale Raleigh signal from horizon-scale scattering at recombination.

\section{Acknowledgments}
I thank Rishi Khatri, Kris Sigurdson, Andrew Jaffe, Anthony Challinor and Duncan Hanson for discussion, \planck\ colleagues for bringing Rayleigh scattering to my attention, and Guido Pettinari for finding an error in the $\nu^8$ factor and a helpful comparison of numerical results.
I acknowledge support from the Science and Technology Facilities Council [grant number ST/I000976/1].
\\
\\
{\bf Notes:}
Since publication in JCAP this version corrects the numerical factor in the $\nu^8$ term of Eq.\eqref{series}, and a $0.7\%$ error in the numerical value of $\nueff$ used for numerical calculations, leading to a small quantitative change in results that does not affect any conclusions.
After submission I also became aware of Rayleigh scattering calculations by E. Alipour, K. Sigurson, and C. Hirata (in preparation), with results for the CMB polarization power spectra similar to those reported here.

%\bibliography{../../Latex/antony,../../Latex/cosmomc}

\begin{thebibliography}{26}
\expandafter\ifx\csname natexlab\endcsname\relax\def\natexlab#1{#1}\fi
\expandafter\ifx\csname bibnamefont\endcsname\relax
  \def\bibnamefont#1{#1}\fi
\expandafter\ifx\csname bibfnamefont\endcsname\relax
  \def\bibfnamefont#1{#1}\fi
\expandafter\ifx\csname citenamefont\endcsname\relax
  \def\citenamefont#1{#1}\fi
\expandafter\ifx\csname url\endcsname\relax
  \def\url#1{\texttt{#1}}\fi
\expandafter\ifx\csname urlprefix\endcsname\relax\def\urlprefix{URL }\fi
\providecommand{\bibinfo}[2]{#2}
\providecommand{\eprint}[2][]{\url{#2}}

\bibitem[{\citenamefont{Takahara and Sasaki}(1991)}]{Takahara91}
\bibinfo{author}{\bibfnamefont{F.}~\bibnamefont{Takahara}} \bibnamefont{and}
  \bibinfo{author}{\bibfnamefont{S.}~\bibnamefont{Sasaki}},
  \bibinfo{journal}{Prog. Theor. Phys} \textbf{\bibinfo{volume}{86}},
  \bibinfo{pages}{1021} (\bibinfo{year}{1991}).

\bibitem[{\citenamefont{Yu et~al.}(2001)\citenamefont{Yu, Spergel, and
  Ostriker}}]{Yu:2001gw}
\bibinfo{author}{\bibfnamefont{Q.-J.} \bibnamefont{Yu}},
  \bibinfo{author}{\bibfnamefont{D.~N.} \bibnamefont{Spergel}},
  \bibnamefont{and} \bibinfo{author}{\bibfnamefont{J.~P.}
  \bibnamefont{Ostriker}}, \bibinfo{journal}{Astrophys.J.}
  \textbf{\bibinfo{volume}{558}}, \bibinfo{pages}{23} (\bibinfo{year}{2001}),
  \eprint{astro-ph/0103149}.

\bibitem[{\citenamefont{{Lee}}(2005)}]{HeeWon05}
\bibinfo{author}{\bibfnamefont{H.-W.} \bibnamefont{{Lee}}},
  \bibinfo{journal}{\mnras} \textbf{\bibinfo{volume}{358}},
  \bibinfo{pages}{1472} (\bibinfo{year}{2005}).

\bibitem[{\citenamefont{Tarafdar and Vardya}(1969)}]{Tarafdar69}
\bibinfo{author}{\bibfnamefont{S.}~\bibnamefont{Tarafdar}} \bibnamefont{and}
  \bibinfo{author}{\bibfnamefont{M.~S.} \bibnamefont{Vardya}},
  \bibinfo{journal}{MNRAS} \textbf{\bibinfo{volume}{145}}, \bibinfo{pages}{171}
  (\bibinfo{year}{1969}).

\bibitem[{\citenamefont{Seager et~al.}(2000)\citenamefont{Seager, Sasselov, and
  Scott}}]{Seager:1999km}
\bibinfo{author}{\bibfnamefont{S.}~\bibnamefont{Seager}},
  \bibinfo{author}{\bibfnamefont{D.~D.} \bibnamefont{Sasselov}},
  \bibnamefont{and} \bibinfo{author}{\bibfnamefont{D.}~\bibnamefont{Scott}},
  \bibinfo{journal}{Astrophys. J. Suppl.} \textbf{\bibinfo{volume}{128}},
  \bibinfo{pages}{407} (\bibinfo{year}{2000}), \eprint{astro-ph/9912182}.

\bibitem[{\citenamefont{Ali-Haimoud and Hirata}(2011)}]{AliHaimoud:2010dx}
\bibinfo{author}{\bibfnamefont{Y.}~\bibnamefont{Ali-Haimoud}} \bibnamefont{and}
  \bibinfo{author}{\bibfnamefont{C.~M.} \bibnamefont{Hirata}},
  \bibinfo{journal}{Phys.Rev.} \textbf{\bibinfo{volume}{D83}},
  \bibinfo{pages}{043513} (\bibinfo{year}{2011}), \eprint{1011.3758}.

\bibitem[{\citenamefont{Chluba and Thomas}(2011)}]{Chluba:2010ca}
\bibinfo{author}{\bibfnamefont{J.}~\bibnamefont{Chluba}} \bibnamefont{and}
  \bibinfo{author}{\bibfnamefont{R.~M.} \bibnamefont{Thomas}},
  \bibinfo{journal}{\mnras} \textbf{\bibinfo{volume}{412}},
  \bibinfo{pages}{748} (\bibinfo{year}{2011}), \eprint{1010.3631}.

\bibitem[{\citenamefont{Lewis and Challinor}(2007)}]{Lewis:2007kz}
\bibinfo{author}{\bibfnamefont{A.}~\bibnamefont{Lewis}} \bibnamefont{and}
  \bibinfo{author}{\bibfnamefont{A.}~\bibnamefont{Challinor}},
  \bibinfo{journal}{Phys. Rev.} \textbf{\bibinfo{volume}{D76}},
  \bibinfo{pages}{083005} (\bibinfo{year}{2007}), \eprint{astro-ph/0702600}.

\bibitem[{\citenamefont{Basu et~al.}(2004)\citenamefont{Basu,
  Hernandez-Monteagudo, and Sunyaev}}]{Basu:2003th}
\bibinfo{author}{\bibfnamefont{K.}~\bibnamefont{Basu}},
  \bibinfo{author}{\bibfnamefont{C.}~\bibnamefont{Hernandez-Monteagudo}},
  \bibnamefont{and} \bibinfo{author}{\bibfnamefont{R.}~\bibnamefont{Sunyaev}},
  \bibinfo{journal}{Astron.Astrophys.} \textbf{\bibinfo{volume}{416}},
  \bibinfo{pages}{447} (\bibinfo{year}{2004}), \eprint{astro-ph/0311620}.

\bibitem[{\citenamefont{Rubino-Martin et~al.}(2005)\citenamefont{Rubino-Martin,
  Hernandez-Monteagudo, and Sunyaev}}]{RubinoMartin:2005dm}
\bibinfo{author}{\bibfnamefont{J.}~\bibnamefont{Rubino-Martin}},
  \bibinfo{author}{\bibfnamefont{C.}~\bibnamefont{Hernandez-Monteagudo}},
  \bibnamefont{and} \bibinfo{author}{\bibfnamefont{R.}~\bibnamefont{Sunyaev}},
  \bibinfo{journal}{Astron.Astrophys.}  (\bibinfo{year}{2005}),
  \eprint{astro-ph/0502571}.

\bibitem[{\citenamefont{Hernandez-Monteagudo
  et~al.}(2006)\citenamefont{Hernandez-Monteagudo, Rubino-Martin, and
  Sunyaev}}]{Hernandez-Monteagudo:2006ar}
\bibinfo{author}{\bibfnamefont{C.}~\bibnamefont{Hernandez-Monteagudo}},
  \bibinfo{author}{\bibfnamefont{J.~A.} \bibnamefont{Rubino-Martin}},
  \bibnamefont{and} \bibinfo{author}{\bibfnamefont{R.~A.}
  \bibnamefont{Sunyaev}} (\bibinfo{year}{2006}), \eprint{astro-ph/0611497}.

\bibitem[{\citenamefont{Schleicher et~al.}(2008)\citenamefont{Schleicher,
  Galli, Palla, Camenzind, Klessen et~al.}}]{Schleicher:2008ji}
\bibinfo{author}{\bibfnamefont{D.}~\bibnamefont{Schleicher}},
  \bibinfo{author}{\bibfnamefont{D.}~\bibnamefont{Galli}},
  \bibinfo{author}{\bibfnamefont{F.}~\bibnamefont{Palla}},
  \bibinfo{author}{\bibfnamefont{M.}~\bibnamefont{Camenzind}},
  \bibinfo{author}{\bibfnamefont{R.}~\bibnamefont{Klessen}},
  \bibnamefont{et~al.}, \bibinfo{journal}{\aap} \textbf{\bibinfo{volume}{490}},
  \bibinfo{pages}{521} (\bibinfo{year}{2008}), \eprint{0803.3987}.

\bibitem[{\citenamefont{Sunyaev and Khatri}(2013)}]{Sunyaev:2013aoa}
\bibinfo{author}{\bibfnamefont{R.~A.} \bibnamefont{Sunyaev}} \bibnamefont{and}
  \bibinfo{author}{\bibfnamefont{R.}~\bibnamefont{Khatri}},
  \bibinfo{journal}{Int.J.Mod.Phys.} \textbf{\bibinfo{volume}{D22}},
  \bibinfo{pages}{1330014} (\bibinfo{year}{2013}), \eprint{1302.6553}.

\bibitem[{\citenamefont{Lewis et~al.}(2000)\citenamefont{Lewis, Challinor, and
  Lasenby}}]{Lewis:1999bs}
\bibinfo{author}{\bibfnamefont{A.}~\bibnamefont{Lewis}},
  \bibinfo{author}{\bibfnamefont{A.}~\bibnamefont{Challinor}},
  \bibnamefont{and} \bibinfo{author}{\bibfnamefont{A.}~\bibnamefont{Lasenby}},
  \bibinfo{journal}{Astrophys. J.} \textbf{\bibinfo{volume}{538}},
  \bibinfo{pages}{473} (\bibinfo{year}{2000}), \eprint{astro-ph/9911177}.

\bibitem[{\citenamefont{Hannestad}(2001)}]{Hannestad:2000fy}
\bibinfo{author}{\bibfnamefont{S.}~\bibnamefont{Hannestad}},
  \bibinfo{journal}{New Astron.} \textbf{\bibinfo{volume}{6}},
  \bibinfo{pages}{17} (\bibinfo{year}{2001}), \eprint{astro-ph/0008452}.

\bibitem[{\citenamefont{Hernandez-Monteagudo and
  Sunyaev}(2005)}]{HernandezMonteagudo:2004xg}
\bibinfo{author}{\bibfnamefont{C.}~\bibnamefont{Hernandez-Monteagudo}}
  \bibnamefont{and} \bibinfo{author}{\bibfnamefont{R.}~\bibnamefont{Sunyaev}},
  \bibinfo{journal}{MNRAS} \textbf{\bibinfo{volume}{359}}, \bibinfo{pages}{597}
  (\bibinfo{year}{2005}), \eprint{astro-ph/0405487}.

\bibitem[{\citenamefont{{Safari} et~al.}(2012)\citenamefont{{Safari}, {Amaro},
  {Fritzsche}, {Santos}, {Tashenov}, and {Fratini}}}]{Safari12}
\bibinfo{author}{\bibfnamefont{L.}~\bibnamefont{{Safari}}},
  \bibinfo{author}{\bibfnamefont{P.}~\bibnamefont{{Amaro}}},
  \bibinfo{author}{\bibfnamefont{S.}~\bibnamefont{{Fritzsche}}},
  \bibinfo{author}{\bibfnamefont{J.~P.} \bibnamefont{{Santos}}},
  \bibinfo{author}{\bibfnamefont{S.}~\bibnamefont{{Tashenov}}},
  \bibnamefont{and}
  \bibinfo{author}{\bibfnamefont{F.}~\bibnamefont{{Fratini}}},
  \bibinfo{journal}{\pra} \textbf{\bibinfo{volume}{86}}, \bibinfo{eid}{043405}
  (\bibinfo{year}{2012}), \eprint{1208.3082}.

\bibitem[{\citenamefont{Ma and Bertschinger}(1995)}]{Ma:1995ey}
\bibinfo{author}{\bibfnamefont{C.-P.} \bibnamefont{Ma}} \bibnamefont{and}
  \bibinfo{author}{\bibfnamefont{E.}~\bibnamefont{Bertschinger}},
  \bibinfo{journal}{Astrophys. J.} \textbf{\bibinfo{volume}{455}},
  \bibinfo{pages}{7} (\bibinfo{year}{1995}), \eprint{astro-ph/9506072}.

\bibitem[{\citenamefont{Hu et~al.}(1998)\citenamefont{Hu, Seljak, White, and
  Zaldarriaga}}]{Hu:1998mn}
\bibinfo{author}{\bibfnamefont{W.}~\bibnamefont{Hu}},
  \bibinfo{author}{\bibfnamefont{U.}~\bibnamefont{Seljak}},
  \bibinfo{author}{\bibfnamefont{M.~J.} \bibnamefont{White}}, \bibnamefont{and}
  \bibinfo{author}{\bibfnamefont{M.}~\bibnamefont{Zaldarriaga}},
  \bibinfo{journal}{Phys. Rev.} \textbf{\bibinfo{volume}{D57}},
  \bibinfo{pages}{3290} (\bibinfo{year}{1998}), \eprint{astro-ph/9709066}.

\bibitem[{\citenamefont{Challinor}(2000)}]{Challinor:2000as}
\bibinfo{author}{\bibfnamefont{A.}~\bibnamefont{Challinor}},
  \bibinfo{journal}{Phys. Rev.} \textbf{\bibinfo{volume}{D62}},
  \bibinfo{pages}{043004} (\bibinfo{year}{2000}), \eprint{astro-ph/9911481}.

\bibitem[{\citenamefont{Lewis and Challinor}(2006)}]{Lewis:2006fu}
\bibinfo{author}{\bibfnamefont{A.}~\bibnamefont{Lewis}} \bibnamefont{and}
  \bibinfo{author}{\bibfnamefont{A.}~\bibnamefont{Challinor}},
  \bibinfo{journal}{Phys. Rept.} \textbf{\bibinfo{volume}{429}},
  \bibinfo{pages}{1} (\bibinfo{year}{2006}), \eprint{astro-ph/0601594}.

\bibitem[{\citenamefont{Dunkley et~al.}(2008)\citenamefont{Dunkley, Amblard,
  Baccigalupi, Betoule, Chuss et~al.}}]{Dunkley:2008am}
\bibinfo{author}{\bibfnamefont{J.}~\bibnamefont{Dunkley}},
  \bibinfo{author}{\bibfnamefont{A.}~\bibnamefont{Amblard}},
  \bibinfo{author}{\bibfnamefont{C.}~\bibnamefont{Baccigalupi}},
  \bibinfo{author}{\bibfnamefont{M.}~\bibnamefont{Betoule}},
  \bibinfo{author}{\bibfnamefont{D.}~\bibnamefont{Chuss}}, \bibnamefont{et~al.}
  (\bibinfo{year}{2008}), \eprint{0811.3915}.

\bibitem[{\citenamefont{Kogut et~al.}(2011)\citenamefont{Kogut, Fixsen, Chuss,
  Dotson, Dwek et~al.}}]{Kogut:2011xw}
\bibinfo{author}{\bibfnamefont{A.}~\bibnamefont{Kogut}},
  \bibinfo{author}{\bibfnamefont{D.}~\bibnamefont{Fixsen}},
  \bibinfo{author}{\bibfnamefont{D.}~\bibnamefont{Chuss}},
  \bibinfo{author}{\bibfnamefont{J.}~\bibnamefont{Dotson}},
  \bibinfo{author}{\bibfnamefont{E.}~\bibnamefont{Dwek}}, \bibnamefont{et~al.},
  \bibinfo{journal}{JCAP} \textbf{\bibinfo{volume}{1107}}, \bibinfo{pages}{025}
  (\bibinfo{year}{2011}), \eprint{1105.2044}.

\bibitem[{\citenamefont{Andre et~al.}(2013)}]{Andre:2013afa}
\bibinfo{author}{\bibfnamefont{P.}~\bibnamefont{Andre}} \bibnamefont{et~al.}
  (\bibinfo{collaboration}{PRISM Collaboration}) (\bibinfo{year}{2013}),
  \eprint{1306.2259}.

\bibitem[{\citenamefont{Ade et~al.}(2011)}]{Ade:2011ap}
\bibinfo{author}{\bibfnamefont{P.}~\bibnamefont{Ade}} \bibnamefont{et~al.}
  (\bibinfo{collaboration}{Planck Collaboration}),
  \bibinfo{journal}{Astron.Astrophys.} \textbf{\bibinfo{volume}{536}},
  \bibinfo{pages}{A18} (\bibinfo{year}{2011}), \eprint{1101.2028}.

\bibitem[{\citenamefont{Bouchet et~al.}(2011)}]{Bouchet:2011ck}
\bibinfo{author}{\bibfnamefont{F.}~\bibnamefont{Bouchet}} \bibnamefont{et~al.}
  (\bibinfo{collaboration}{COrE Collaboration}) (\bibinfo{year}{2011}),
  \eprint{1102.2181}.

\end{thebibliography}
\providecommand{\aj}{Astron. J. }\providecommand{\apj}{Astrophys. J.
  }\providecommand{\apjl}{Astrophys. J.
  }\providecommand{\mnras}{MNRAS}\providecommand{\aap}{Astron. Astrophys.}

\end{document}